\documentclass{article}
\usepackage{amssymb}
\newtheorem{proposition}{Proposition}
\newtheorem{theorem}{Theorem}
\newtheorem{lemma}{Lemma}
\newtheorem{corollary}{Corollary}

\renewcommand{\theequation}{\thesection.\arabic{equation}}

\newcommand{\bfp}{{\bf p}}

\newcommand{\bfq}{{\bf q}}
\newcommand{\bfx}{{\bf x}}

\newcommand{\bfa}{{\bf a}}

\newcommand{\bfe}{{\bf e}}
\newcommand{\bfalpha}{\mbox{\boldmath $ \alpha$}}

\newcommand{\bfsigma}{\mbox{\boldmath $ \sigma$}}

\newcommand{\bfnabla}{\mbox{\boldmath $ \nabla$}}
\newcommand{\bfxi}{\mbox{\boldmath $ \xi$}}

\newcommand{\bfpi}{\mbox{\boldmath $ \pi$}}

\newcommand{\cA}{{\mathcal{A}}}

\newcommand{\cO}{{\mathcal{O}}}
\newcommand{\cD}{{\mathcal{D}}}
\newcommand{\ninfi}{{n \rightarrow \infty}}
\newcommand{\nin}{{n \in {\Bbb N}}}

\newenvironment{proof}{{\noindent\it Proof. $\;$}}{\hspace*{\fill} $\square$}

\begin{document}

\begin{center}
{\Large\bf On the essential spectrum of the

Jansen-Hess 
operator for two-electron ions}
\end{center} 

\vspace{0.5cm}

\begin{center}
{\large D.~H.~Jakubassa-Amundsen  }

Mathematics Institute,
University of Munich\\ Theresienstr. 39, 80333
Munich, Germany  
\end{center}

\vspace{2cm}

\begin{abstract}
Based on the HVZ theorem and dilation analyticity 
of the pseudorelativistic no-pair Jansen-Hess operator, it is shown that for subcritical potential strength ($Z\leq 90)$
the singular continuous spectrum is absent.
The bound is slightly higher ($Z\leq 102$) for the Brown-Ravenhall operator whose eigenvalues $\lambda$
are, by the virial theorem, confined to $\lambda <2m$ if $Z\leq 50.$

\end{abstract}

\newpage

\section{Introduction}

We consider two interacting electrons of mass $m$ in a central Coulomb field,
generated by a point nucleus of charge number $Z$ which is fixed at the origin.
The Jansen-Hess operator  that is used for the description of this system, results from a block-diagonalization of
the Coulomb-Dirac operator up to second order in the fine structure constant
 $e^2\approx 1/137.04$ \cite{DK,Jaku3}.
Convergence of this type of expansion has recently been proven for
$Z<52$ \cite{SS,HS}, and numerical higher-order investigations have established
the Jansen-Hess operator as a very good approximation (see e.g. \cite{RW}).

Based on the work of Lewis, Siedentop and Vugalter \cite{LSV}
the essential spectrum of the two-particle Jansen-Hess operator $h^{(2)}$ was localized in $[\Sigma_0,\infty)$ with $ \Sigma_0-m$ being the ground-state energy of the one-electron ion \cite{Jaku2}.
A more detailed information on the essential spectrum exists only for the single-particle Jansen-Hess operator, for which, in case of sufficiently small central potential strength $\gamma$, 
the absence of the singular continuous spectrum $\sigma_{sc}$ and of embedded eigenvalues was proven \cite{Jaku1}.
These results were obtained with the help of scaling properties and dilation analyticity of this operator, combined with the virial theorem,
methods which, initiated by Aguilar and Combes, are well-known from the analysis of the Schr\"{o}dinger operator \cite{AC},\cite[p.231]{Reed4} 
and of the single-particle Brown-Ravenhall operator. For the latter operator,
the absence of $\sigma_{sc}$ as well as of embedded eigenvalues in $[m,\infty)$ was proven for all $\gamma < \gamma_{BR}=2(\frac{2}{\pi}+\frac{\pi}{2})^{-1},\;\;\gamma_{BR}$ being the maximum value for which this operator
is bounded from below \cite{EPS,BE,Jaku1}.
For more than one electron the absence of $\sigma_{sc}$ in the Schr\"{o}dinger case was shown along the same lines \cite{BC,Simon}, the basic ingredient (apart
from the dilation analyticity of the operator) being the relative compactness of the Schr\"{o}dinger potential with respect to the kinetic energy operator.
Such a compactness property does not exist for Dirac-type operators.
For the determination of the spectral properties of $h^{(2)}$ ingredients of complex analysis are used instead to prove a two-particle HVZ-type theorem for non-selfadjoint operators which depend on a complex parameter $\theta$, forming an analytic family and being self-adjoint for real $\theta$ (Proposition 1, section 3).
With this HVZ theorem at hand, the dilation analytic method of Balslev and Combes \cite{BC}
can be used to prove the absence of $\sigma_{sc}$ (Theorem 1, section 3).
Concerning eigenvalues embedded in the essential  spectrum, the virial theorem is formulated for the two-particle operator,
and a modification of the proof by Balinsky and Evans \cite{BE} is tested on $h^{BR}$ to show the absence of eigenvalues in $[2m,\infty)$ (Proposition 2, section 5).

Let us now define our operators in question.
The two-particle pseudorelativistic no-pair Jansen-Hess operator, acting in the Hilbert space $\cA(L_2({\Bbb R}^3)\otimes {\Bbb C}^2)^2$ where $\cA$ denotes antisymmetrization with
respect to particle exchange, is given (in relativistic units, $\hbar = c =1$) by \cite{Jaku3}
\begin{equation}\label{1.1}
h^{(2)}\;=\;h^{BR}\;+\;\sum_{k=1}^2 b_{2m}^{(k)}\;+\;c^{(12)}.
\end{equation}
The term up to first order in $e^2$ 
 is the (two-particle) Brown-Ravenhall operator \cite{BR,EPS,Jaku2a}
$$ h^{BR}\;=\; \sum_{k=1}^2 \left( T^{(k)} + b_{1m}^{(k)}\right) \;+\;v^{(12)},$$
\begin{equation}\label{1.2}
T^{(k)}:=\; E_{p_k}:=\;\sqrt{p_k^2+m^2},\qquad 
b_{1m}^{(k)} \;\sim\; -P_0^{(12)}\,U_0^{(k)}\;\frac{\gamma}{x_k}\;U_0^{(k)-1}\,P_0^{(12)},
\end{equation}
$$ v^{(12)}\;\sim\; P_0^{(12)}\,U_0^{(1)} U_0^{(2)}\;\frac{e^2}{|\bfx_1-\bfx_2|}\;(U_0^{(1)} U_0^{(2)})^{-1}\,P_0^{(12)},$$
where the index $m$ refers to the particle mass,
 $\bfp_k=-i\bfnabla_k$ is the momentum and $\bfx_k$ (with $x_k:= |\bfx_k|$) the location of particle $k$ relative to the origin. $\gamma = Ze^2$ is the central field strength, and $v^{(12)}$ the electron-electron interaction. $U_0^{(k)}$ denotes the unitary Foldy-Wouthuysen transformation,
$$U_0^{(k)}\;=\;A(p_k)\;+\;\beta^{(k)} \bfalpha^{(k)} \bfp_k g(p_k),$$
\begin{equation}\label{1.3}
A(p):=\;\left( \frac{E_p+m}{2E_p}\right)^\frac12,\qquad g(p):=\; \frac{1}{\sqrt{2 E_p(E_p+m)}}
\end{equation}
and the inverse $U_0^{(k)-1}=U_0^{(k)\ast}=A(p_k)+ \bfalpha^{(k)} \bfp_kg(p_k)\beta^{(k)}$ with $\bfalpha^{(k)},\;\beta^{(k)}$ Dirac matrices \cite{Th}.
Finally, $P_0^{(12)} = P_0^{(1)} P_0^{(2)}$ where $P_0^{(k)}:= \frac{1+\beta^{(k)}}{2}$ projects onto the upper two components of the four-spinor of particle $k$ (hence reducing the four-spinor space
to a two-spinor space).

The remaining potentials in (\ref{1.1}) which are of second order in the fine structure constant consist of the single-particle contributions
$$b_{2m}^{(k)}\;\sim\; P_0^{(12)}\,U_0^{(k)}\;\frac{\gamma^2}{8\pi^2}\left\{ \frac{1}{x_k}\,(1-\tilde{D}_0^{(k)})\,V_{10,m}^{(k)}\;+\;h.c.\right\}\;U_0^{(k)-1}\,P_0^{(12)},\quad k=1,2,$$
\begin{equation}\label{1.4}
\tilde{D}_0^{(k)}:=\;\frac{\bfalpha^{(k)} \bfp_k+\beta^{(k)}m}{E_{p_k}},\qquad V_{10,m}^{(k)}:=\;2\pi^2\int_0^\infty dt\;e^{-tE_{p_k}}\,\frac{1}{x_k}\,e^{-tE_{p_k}},
\end{equation}
where $\tilde{D}_0^{(k)}$ has norm unity, $V_{10,m}^{(k)}$ is bounded and $h.c.$ stands for hermitean conjugate (such that $b_{2m}^{(k)}$ is a symmetric operator).
The two-particle interaction is given by
$$c^{(12)}\sim P_0^{(12)}U_0^{(1)} U_0^{(2)}\frac12 \sum_{k=1}^2 \left\{ \frac{e^2}{|\bfx_1-\bfx_2|}\,(1-\tilde{D}_0^{(k)})\,F_0^{(k)}+h.c.\!\right\}\!\left( U_0^{(1)} U_0^{(2)}\right)^{-1}\!\!\!P_0^{(12)}\!\!,$$
\begin{equation}\label{1.5}
F_0^{(k)}:=\;-\frac{\gamma}{2}\int_0^\infty dt\;e^{-tE_{p_k}}\left( \frac{1}{x_k}\,-\,\tilde{D}_0^{(k)}\,\frac{1}{x_k}\,\tilde{D}_0^{(k)}\right)\;e^{-tE_{p_k}}.
\end{equation}
For later use, we also provide the kernel of the bounded operator $F_0^{(k)}$ in momentum space,
\begin{equation}\label{1.6}
k_{F_0^{(k)}}(\bfp,\bfp')\;=\;-\frac{\gamma}{(2\pi)^2}\;\frac{1}{|\bfp-\bfp'|^2}\;\frac{1}{E_{p}+E_{p'}}\left( 1\,-\,\tilde{D}_0^{(k)}(\bfp)\,\tilde{D}_0^{(k)}(\bfp')\right).
\end{equation}
The notation l.h.s. $\sim$ r.h.s. in (\ref{1.2}) -- (\ref{1.5})  means that the l.h.s. is defined by the nontrivial part (i.e. the upper block) of the r.h.s.
(see e.g. \cite{EPS,Jaku3}).

$h^{(2)}$ is a well-defined operator in the form sense for $\gamma < 0.98\;$ (which follows from the form boundedness of the 
Jansen-Hess potential with respect to the kinetic energy with relative bound less than one;
see section 2 for the improvement of the bound 0.89 given in \cite{Jaku3}), 
and is self-adjoint by means of its Friedrichs extension.

\section{Dilation analyticity}
\setcounter{equation}{0}

For a one-particle function $\varphi \in L_2({\Bbb R}^3)\otimes {\Bbb C}^2$ and $\theta:= e^\xi \in {\Bbb R}_+$ we define the unitary group of dilation operators $d_\theta$ by means of \cite{AC}
\begin{equation}\label{2.1}
d_\theta \varphi(\bfp):=\;\theta^{-3/2}\,\varphi(\bfp/\theta)
\end{equation}
with the property
\begin{equation}\label{2.2}
d_{\theta_{1}} d_{\theta_2}\varphi(\bfp)\,=\,(\theta_{1}\theta_2)^{-3/2}\,\varphi(\bfp/\theta_{1}\theta_2)\;=\;d_\theta\;\varphi(\bfp)
\end{equation}
where $\theta:= \theta_{1} \theta_2=e^{\xi_1+\xi_2}.\;$
For a two-particle function $\psi \in \cA(L_2({\Bbb R}^3)\otimes {\Bbb C}^2)^2$ we have
$d_\theta \,\psi(\bfp_1,\bfp_2)\,=\theta^{-3}\psi(\bfp_1/\theta,\bfp_2/\theta).$

Let $\cO_\theta:= d_\theta \cO d_\theta^{-1}$ be the dilated operator $\cO\;$ (e.g. $h_\theta^{(2)}:= d_\theta h^{(2)} d_\theta^{-1}).\;$ From the explicit structure of the summands of $h^{(2)}$ in momentum space one derives
the following scaling properties, using the form invariance $(\psi, h^{(2)} \psi)\,=
(d_\theta \psi, h_\theta^{(2)} d_\theta\psi)\;$ for $\psi \in \cA(H_{1/2}({\Bbb R}^3)\otimes {\Bbb C}^2)^2$,
the form domain of $h^{(2)}$ (see \cite{EPS,Jaku1}, \cite[p.42,73]{thesis}),
\begin{equation}\label{2.3} 
T_\theta^{(k)}(m)\;=\;\sqrt{p_k^2/\theta^2+m^2}\;=\;\frac{1}{\theta}\;
\sqrt{p_k^2+m^2\theta^2}\;=\;\frac{1}{\theta}\;T^{(k)}(m\cdot \theta)
\end{equation}
$$h_\theta^{BR}(m)\;=\; \frac{1}{\theta}\;h^{BR}(m \cdot \theta),\qquad h_\theta^{(2)}(m)\;=\;\frac{1}{\theta}\;h^{(2)}(m \cdot \theta)$$
where we have indicated explicitly the mass dependence of the operators.

Let us extend $\theta$ to a domain $\cD$ in the complex plane,
\begin{equation}\label{2.4}
\cD:=\; \{ \theta \in {\Bbb C}:\;\theta=e^\xi,\;|\xi|<\xi_0\},
\end{equation}
with $0<\xi_0<\frac12$ to be fixed later. The definition of the dilated operators with the scaling properties (\ref{2.3}) is readily extended to $\theta \in \cD.$

In order to establish the existence of $h_\theta^{(2)}$ for $\theta \in \cD$ as a form sum one has for $T_\theta:= T_\theta^{(1)} + T_\theta^{(2)}$ to
assure the $|T_\theta|$-form boundedness of the potential of $h_\theta^{(2)}$ with relative bound smaller than one. For the single-particle contributions this was shown earlier for potential strength $\gamma < 1.006$ \cite{Jaku1}.

Let us start by noting that the $m$-dependent factors appearing in the potential terms of $h^{(2)}$ are all of the form $E_p^\lambda,\;(E_p+m)^\lambda,\;\;\lambda \in {\Bbb R},\;$ as well as  $\frac{1}{E_{p}+E_{p'}}$
(see e.g. (\ref{1.3}), (\ref{1.6})). This assures that $h^{(2)}\psi$ is an analytic function of $m$
for $m \neq 0.$ 

For $\theta \in \cD$ we basically have to replace $m$ by $m \cdot \theta$. We can use estimates of the type \cite{Jaku1}
$$1-\xi_0\;\leq\;\left| \frac{1}{\theta}\right|\;\leq\; 1+2\xi_0$$
\begin{equation}\label{2.5}
(1-\xi_0)\;E_p\;\leq\; |E_\theta(p)|\;\leq \;(1+2\xi_0)\;E_p
\end{equation}
where $E_\theta(p): =\sqrt{p^2+m^2\theta^2}\,$.
From these relations one derives the relative boundedness of the following dilated operators with respect
to those for $\theta=1$,
$$|A_\theta(p)|^2\;\leq\; \frac{1+2\xi_0}{1-\xi_0}\;A^2(p)$$
\begin{equation}\label{2.5a}
|\frac{p}{\theta}\;g_\theta(p)|^2\;\leq\; \frac{1}{(1-\xi_0)^4}\;p^2\;g^2(p)
\end{equation}
$$\left| \frac{1}{E_\theta(p)+E_\theta(p')}\right|\;
\leq\; \frac{1}{(1-\xi_0)^3}\;\frac{1}{E_p+E_{p'}}.$$
As a consequence, the dilated Foldy-Wouthuysen transformation is bounded,
$|U_\theta^{(k)}|\,\\ \leq\, |A_\theta(p_k)|\,+|\frac{p_k}{\theta}\,g_\theta(p_k)|\,\leq\tilde{c},\;$ and also
$|\tilde{D}_{0,\theta}^{(k)}|\,\leq\,\frac{1}{|E_\theta(p_k)|}\,(p_k+m|\theta|)\,\leq \tilde{c}$ with some constant $\tilde{c}$.

In order to show the relative form boundedness of $h_\theta^{(2)}$, we write $h^{(2)}=T+W$ and introduce the respective massless ($m=0$) operators $T_0 = p_1+p_2$ and $W_0$,
\begin{equation}\label{2.6}
|(\psi, W_\theta\psi)|\;\leq\; |\frac{1}{\theta}\,(\psi, W_0\psi)|\;+\;|(\psi, \left( W_\theta -\frac{1}{\theta}\,W_0\right)\psi)|.
\end{equation}
The form boundedness of $W_0$ with respect to $T_0$
 follows from the previous single-particle \cite{BSS} and two-particle \cite{Jaku3} $m=0$ estimates. 
For the single-particle contributions we profit from \cite{BSS}
$(\psi, (p_k+b_1^{(k)}+b_2^{(k)})\,\psi)\geq \,(1-\frac{\gamma}{\gamma_{BR}}+d\gamma^2)(\psi,p_k\psi)\;$ together
with \cite{Jaku1} $b_1^{(k)}+b_2^{(k)} <0$ for $\gamma \leq \frac{4}{\pi}.\;$ Note that (e.g. for
 $k=1$) $\;\psi=\psi_{\bfx_2}(\bfx_1)$ acts as a one-particle function depending parametrically on the coordinates of the second particle.
For the two-particle terms, use is made of
$(U_0^{(k)\ast} \psi_0, p_kU_0^{(k)\ast}\,\psi_0)\,=(\psi,p_k\,\psi)$
where $\psi_0:= {\psi \choose 0}$ denotes a two-particle spinor whose lower
components are zero by the action of $P_0^{(12)},$ showing that the four-spinor estimates from \cite{Jaku3} are applicable. Thus,
$$|(\psi, W_0\,\psi)|\;\leq\; \sum_{k=1}^2 |(\psi,(b_1^{(k)}+b_2^{(k)})\;\psi)|\;+\;|(\psi,c_0^{(12)}\;\psi)|\;+\;|(\psi,v_0^{(12)}\;\psi)|$$
\begin{equation}\label{2.6a}
\leq\;\left( \frac{\gamma}{\gamma_{BR}}-d\gamma^2+\gamma\;\frac{e^2\pi^2}{4}\;+\;\frac{e^2}{2\gamma_{BR}}\right)\;(\psi,T_0\;\psi)\;=:\;\tilde{c}_0\;(\psi,T_0\;\psi),
\end{equation}
where
 $\gamma_{BR}\approx 0.906$ and $d=\frac{1}{8}\left(\frac{\pi}{2}- \frac{2}{\pi}\right)^2.\;$

For the proof of the form boundedness with respect to $|T_\theta|$, we can estimate for $|$Im $\xi| <\frac{\pi}{4}$ \cite{Jaku1}
\begin{equation}\label{2.9a}
\mbox{Re }\sqrt{p_k^2+m^2\theta^2}\;\geq\; p_k\,\cos(\mbox{Im }\xi)\;\geq\;p_k\;(1-\xi_0)
\end{equation}
such that
\begin{equation}\label{2.7}
|\theta|\cdot |(\psi, T_\theta^{(k)} \psi)|\;\geq\; |\mbox{Re }(\psi, \sqrt{p_k^2+m^2\theta^2}\,\psi)|\;\geq\;(1-\xi_0)\;(\psi, T_0^{(k)}\psi).
\end{equation}
The uniform boundedness of the single-particle remainder in (\ref{2.6}),
$\frac{1}{|\theta|}\,|(\psi,(b_{1m\cdot \theta}^{(k)}\\-b_1^{(k)})\,\psi)|\,+\,\frac{1}{|\theta|}\,|(\psi,(b_{2m\cdot \theta}^{(k)} -b_2^{(k)})\,\psi)|$ was proven in \cite{Jaku1} based on the respective results for $\theta =1$ \cite{Tix,BSS}.

For the proof of the uniform boundedness of $(\psi,(c^{(12)}(m\cdot \theta)-c_0^{(12)})\,\psi)\;$ and $(\psi,(v^{(12)}(m\cdot \theta)-v_0^{(12)})\,\psi)\;$
we proceed in a similar way. Since $c^{(12)}$ and $v^{(12)}$ are analytic functions of $m$, the mean value theorem can be applied in the form
$|f(m\cdot \theta)-f(0)|\,\leq m(\,|\frac{\partial f}{\partial m}(\tilde{m}_1\cdot \theta)|\,+\,|\frac{\partial f}{\partial m}(\tilde{m}_2 \cdot \theta)|\,)$
with $0\leq \tilde{m}_1,\tilde{m}_2\leq m\;$ (adapted to complex-valued functions \cite{Jaku1}).
The kernel of $v^{(12)}$ is given by $K_{v^{(12)}}(\bfp_1,\bfp_2;\bfp'_1,\bfp'_2):= U_0^{(2)} U_0^{(1)} k_{v^{(12)}} U_0^{(1')\ast} U_0^{(2')\ast}$ with
\begin{equation}\label{2.10a}
k_{v^{(12)}}:=\;\frac{e^2}{2\pi^2}\;\frac{1}{|\bfp_1-\bfp'_1|^2}\;\delta(\bfp'_2-\bfp_2+\bfp'_1-\bfp_1),
\end{equation}
such that one gets
$$\left| (K_{v^{(12)}}(m\cdot \theta)-K_{v_0^{(12)}})(\bfp_1,\bfp_2;\bfp'_1,\bfp'_2)\right|
\;\leq\; m\,k_{v^{(12)}}$$
\begin{equation}\label{2.7a}
\cdot\left(\,\left| \frac{\partial}{\partial m}\left( U_0^{(1)}U_0^{(2)}U_0^{(1')\ast}U_0^{(2')\ast}\right)(\tilde{m}_1\cdot \theta) 
\right|\;+\;(\tilde{m}_1 \mapsto \tilde{m}_2)\;\right),
\end{equation}
where $(\tilde{m}_1 \mapsto \tilde{m}_2)$ means the first term in the second line
of (\ref{2.7a}) repeated with $\tilde{m}_1$ replaced by $\tilde{m}_2$,
and $U_0^{(k')}$ is $U_0^{(k)}$ with $\bfp_k$ replaced by $\bfp'_k.$ 
Further,
$$\left| \frac{\partial}{\partial m}\;(U_0^{(1)} \cdots U_0^{(2')\ast})\,\right|\;\leq\; \left|\frac{\partial U_0^{(1)}}{\partial m}\right|\cdot\, \left| U_0^{(2)}\, U_0^{(1')\ast}U_0^{(2')\ast}\right|\;+\;...$$
\begin{equation}\label{2.7b}
+\;\left| U_0^{(1)}U_0^{(2)}\,U_0^{(1')\ast}\right| \cdot\, \left| \frac{\partial U_0^{(2')\ast}}{\partial m}\right|.
\end{equation}
From the boundedness of $U_0^{(k)}$ and of $\theta$ one gets the estimate (noting that $U_0^{(k)}$ is only a function of $m/p_k=:\xi)$
$$\left| \frac{\partial}{\partial m}\;U_0^{(k)}(\xi \cdot \theta)\right|\;=\;\frac{|\theta|}{p_k}\;\left| \frac{\partial}{\partial (\xi \cdot \theta)}\;U_0^{(k)}(\xi \cdot \theta)\right|$$
\begin{equation}\label{2.7c}
\leq\; \frac{|\theta|}{p_k}\;\frac{c}{1+\xi}\;\leq \;\frac{\tilde{c}}{p_k+m}\;\leq\;\frac{\tilde{c}}{p_k}
\end{equation}
with some constants $c,\tilde{c}$ independent of $m$.
With this estimate the boundedness of $v^{(12)}(m\cdot \theta)-v_0^{(12)}$ is readily shown (see e.g.
\cite{Jaku3} and Appendix A, where a sketch of the boundedness proof for
$c^{(12)}(m\cdot \theta)-c_0^{(12)}$ is given).

Thus we obtain
\begin{equation}\label{2.7d}
 |(\psi, W_\theta \psi)|\,\leq \frac{\tilde{c}_0}{1-\xi_0}\,|(\psi, T_\theta \psi)|\,+C(\psi,\psi)\;
\end{equation}
with $\tilde{c}_0$ from (\ref{2.6a}) and some constant $C$. We have $\tilde{c}_0<1\;$ (and hence also $\frac{\tilde{c}_0}{1-\xi_0}<1\;$ for $\xi_0$ sufficiently small) for $\gamma <0.98\;\;(Z\leq 134).\;$
This holds  for all $\theta\in \cD.$
Besides this $T_0$- and $T_\theta$-form boundedness with $\tilde{c}_0<1$, (\ref{2.3})
assures that for $\psi$ in the form domain of $T_0,\;\;(\psi, h_\theta^{(2)} \psi)$ is an analytic function in $\cD$.
Thus $h_\theta^{(2)}$ 
satisfies the criterions for being a dilation analytic family in the form sense \cite{EPS},\cite[p.20]{Reed4}.

\section{Main theorem and outline of proof}
\setcounter{equation}{0}

The aim of the present work is to prove

\begin{theorem}\label{t1}
Let $h^{(2)}$ be the two-particle Jansen-Hess operator and assume $\gamma\leq 0.66\;\;(Z\leq 90).\;$ Then the singular continuous spectrum is absent,
$$\sigma_{sc}(h^{(2)})\;=\;\emptyset.$$
\end{theorem}

The basic ingredient of the proof is a HVZ-type theorem for nonsymmetric dilation-analytic potentials.

\begin{proposition}\label{p1}
Let $h_\theta^{(2)}\,=\sum\limits_{k=1}^2 (T_\theta^{(k)} +b_{1m,\theta}^{(k)} + b_{2m,\theta}^{(k)})\,+ v_\theta^{(12)}+c_\theta^{(12)}$ be the
dilated two-particle Jansen-Hess operator and let $\theta \in \cD \subset {\Bbb C}.\;$ Let
\begin{equation}\label{3.1}
h_\theta^{(2)}\;=\;T_\theta\,+\,a_{1,\theta}\,+\,r_{1,\theta}
\end{equation}
be the two-cluster decomposition which corresponds to moving particle 1 to infinity.
Then for $\gamma \leq 0.66,$ the essential spectrum of $h_\theta^{(2)}$ is given by
\begin{equation}\label{3.2}
\sigma_{ess}(h_\theta^{(2)})\;=\;\sigma(T_\theta +a_{1,\theta})
\end{equation}
where 
\begin{equation}\label{3.3}
\sigma(T_\theta + a_{1,\theta})\;=\; \sigma(T_\theta^{(1)})\,+\,\sigma(T_\theta^{(2)} + b_{1m,\theta}^{(2)}+b_{2m,\theta}^{(2)})
\end{equation}
and $r_{1,\theta}=b_{1m,\theta}^{(1)}+b_{2m,\theta}^{(1)}+v_\theta^{(12)}+c_\theta^{(12)}.$
\end{proposition}

Starting point of the proof of Theorem \ref{t1} is the invariance of the resolvent form under dilations with $\theta \in \cD \cap {\Bbb R},$
\begin{equation}\label{3.4}
(\psi, \frac{1}{h^{(2)} -z}\,\psi)\;=\;(d_\theta \psi,\frac{1}{h_\theta^{(2)}-z}\;d_\theta \psi)\qquad \mbox{ for } z \in {\Bbb C} \backslash {\Bbb R}.
\end{equation}
Let us restrict ourselves to analytic vectors $\psi \in \cA(N_{\xi_0} \otimes {\Bbb C}^2)^2$
where $N_{\xi_0}:=\{\varphi \in H_{1/2}({\Bbb R}^3):\,d_\theta \varphi $ is analytic in $\cD\}.\;$ For $z \in {\Bbb C}\backslash \sigma(h_\theta^{(2)}),$
the  analyticity of $(h_\theta^{(2)}-z)^{-1}$ and of the function $d_\theta \psi$  allows for the extension of the r.h.s. of (\ref{3.4}) to complex $\theta \in \cD.$
The identity theorem of complex analysis then guarantees the equality (\ref{3.4}) for all $\theta \in \cD.$
Since $N_{\xi_0}$ is dense in $H_{1/2}$ \cite[p.187]{Reed4}, (\ref{3.4}) holds for all
$\psi$ in  $\cA(H_{1/2}({\Bbb R}^3) \otimes {\Bbb C}^2)^2$.

From Proposition \ref{p1} we know that $h_\theta^{(2)}$ has only discrete spectrum $(\sigma_d)$
outside $\sigma(T_\theta +a_{1,\theta})$.

Let us therefore shortly investigate the spectrum of $T_\theta+a_{1,\theta}.$
From the explicit expression $T_\theta^{(1)}=\sqrt{p_1^2/\theta^2+m^2},\;\; p_1 \geq 0,$ it follows that $\sigma(T_\theta^{(1)})=\sigma_{ess}(T_\theta^{(1)})$ is for each $\theta \in \cD$ a curve in the complex plane intersecting ${\Bbb R}$ only in the point $m$ \cite{Wed,Herbst}.

Concerning the spectrum of $b_{m,\theta}^{(2)}:= T_\theta^{(2)}+b_{1m,\theta}^{(2)}+b_{2m,\theta}^{(2)}$, it was shown in \cite{Jaku1} that $\sigma_{ess}(b_{m,\theta}^{(2)})=\sigma_{ess}(T_\theta^{(2)})$ based on the compactness of the difference of the resolvents
of $b_{m,\theta}^{(2)}$ and $T_\theta^{(2)}.$
Thus we get from (\ref{3.3})
$$\sigma(T_\theta+a_{1,\theta})=\{\sqrt{p_1^2/\theta^2+m^2}: p_1 \geq 0\}+\left(\!\! \{\sqrt{p_2^2/\theta^2+m^2}:p_2 \geq 0\}\cup\,\sigma_d(b_{m,\theta}^{(2)})\right)
$$
\begin{equation}\label{3.5}
=\;\sigma_{ess}(T_\theta+a_{1,\theta}),
\end{equation}
which means that $\sigma(T_\theta+a_{1,\theta})$ consists of a system of parallel curves each starting at $m+\lambda_2^{(\theta)}$ for any $\lambda_2^{(\theta)} \in \sigma_d(b_{m,\theta}^{(2)})$, supplied
by an area in the complex plane bounded to the right by such a curve starting at the point $2m$.
(The left boundary is a line starting at $2m$ with $e^{-i\, \mbox{{\scriptsize Im}}\,\xi}{\Bbb R}_+$ as asymptote.)
 The curve $\{\sqrt{p^2/\theta^2+m^2}\,:\,p\geq 0\}$ attached to each $\lambda_2^{(\theta)}$ lies in the closed half plane below (respectively above)
 the real axix, if $\theta=e^\xi$ with Im $\xi >0\;$ (respectively Im $\xi <0),\;$ and its asymptote is $e^{-i\,{\rm Im}\,\xi}\,{\Bbb R}_++\lambda_2^{(\theta)}.$

Any such $\lambda_2^{(\theta)}$ is a discrete eigenvalue of finite multiplicity. Therefore,
since $b_{m,\theta}^{(2)}$ is a dilation analytic operator in $\cD$ it follows from \cite[p.387]{Kato},\cite[p.22]{Reed4} that $\lambda_2^{(\theta)}$  is an analytic function of $\theta$ in $\cD$ (as
long as it remains an isolated eigenvalue).
If $\theta \in {\Bbb R} \,\cap \cD,\;\;\lambda_2^{(\theta)}= \lambda_2^{(1)}\in \sigma_d(b_m^{(2)})$ because $d_\theta$ is unitary for real $\theta$. It then follows from the identity theorem of complex analysis that $\lambda_2^{(\theta)}=\lambda_2^{(1)}$ for all $\theta \in \cD$ \cite{AC}.
Conversely, assume there exists $\tilde{\lambda}_2^{(\theta)} \in \sigma_d(b_{m,\theta}^{(2)})$ in ${\Bbb C}\backslash {\Bbb R}$
(called 'resonance' \cite[p.191]{Reed4}) for a given $\theta \in \cD.$
Then from the group property (\ref{2.2}), a further dilation by any $\tilde{\theta} \in {\Bbb R}$ leaves $\tilde{\lambda}_2^{(\theta)}$ invariant. Thus $\tilde{\lambda}_2^{(\theta)}$ is invariant in the subset of $\cD$
in which it is analytic.

To be specific, let Im $\xi >0.$ 
As a consequence \cite{Wed}, resonances are only possible in the sector bounded
by $\sigma(T_\theta^{(2)})$ and $[m,\infty).$
In particular, no elements of $\sigma_d(b_{m,\theta}^{(2)})$ lie in the upper half plane
(they would be isolated for all $\theta$ with Im $\xi \,\geq 0,$ but such elements have to be real).
Moreover, they can at most accumulate at $m$. (If they did accumulate at some $z_0 \in \sigma(T_\theta^{(2)}) \backslash \{m\}$ then, for $\theta_0=e^{\xi + i\delta}\;\;(\delta >0)\;$
they would, due to their $\theta$-invariance, still accumulate at $z_0 \notin \sigma(T_{\theta_0}^{(2)})$
which is impossible.)
Likewise, real elements of $\sigma_d(b_{m,\theta}^{(2)})$ can only accumulate at $m$. 
Therefore, the intersection set $M_{\Bbb R}:= \sigma(T_\theta +a_{1,\theta}) \cap {\Bbb R}$
consists of $2m$ plus isolated points which can at most accumulate at $2m$.

We note, however, that each of the elements of $M_{\Bbb R}$ can be an accumulation point of $\sigma_d(h_\theta^{(2)}),$ due to Proposition \ref{p1}.
(The nonreal elements of $\sigma_d(h_\theta^{(2)})$ again have to lie in a sector of the lower
half plane, bounded by $\sigma(T_\theta^{(1)})$ and $[m,\infty).)\;$
 From (\ref{3.4}) we get
\begin{equation}\label{3.6}
\lim_{\mbox{{\scriptsize Im}}\,z \rightarrow 0} \left| \mbox{Im }(\psi, \frac{1}{h^{(2)}-z}\;\psi)\right|\;\leq\; \lim_{\mbox{{\scriptsize Im}}\,z \rightarrow 0}
\left| (\psi,\frac{1}{h^{(2)}-z}\;\psi)\right|\;<\;\infty
\end{equation}
for Re $z\notin M_{\Bbb R} \cup \sigma_d(h_\theta^{(2)}),\;$
such that the singular continuous spectrum is absent for ${\Bbb R} \backslash (M_{\Bbb R} \cup \sigma_d(h_\theta^{(2)})$ \cite[p.137]{Reed4}.

To proceed further we follow the argumentation of Balslev and Combes \cite{BC} from the Schr\"{o}dinger case and denote by $\Sigma$ the set of accumulation points of
$M_{\Bbb R} \cup \sigma_d(h_\theta^{(2)}).$ In all other points the spectrum is discrete, and we have no singular continuous spectrum in ${\Bbb R} \backslash \Sigma.$
Let now $\{E_\lambda\}_{\lambda \in {\Bbb R}}$ be the spectral projection of $h^{(2)}$ and let $\psi_{sc}$ be an element of the singular continuous subspace of $\cA(L_2({\Bbb R}^3) \otimes {\Bbb C}^2)^2$.
Then $(\psi_{sc},E_\lambda \psi_{sc})=0$ for all $\lambda \in {\Bbb R}\backslash \Sigma.\;$
Since $(\psi_{sc},E_\lambda \psi_{sc})$ is continuous \cite[p.517]{Kato}
and since $\Sigma$ consists only of isolated points with a possible accumulation point at $2m$,
it follows that $(\psi_{sc},E_\lambda \psi_{sc})=0$ for all $\lambda \in {\Bbb R}.$ Thus $\sigma_{sc}(h^{(2)})=\emptyset.$

\vspace{0.2cm}

\begin{corollary} \label{c1}
For the two-particle Brown-Ravenhall operator $h^{BR}$ we have\\ $\sigma_{sc} (h^{BR}) = \emptyset\;$ if
$\gamma <0.74\;\;(Z< 102).$
\end{corollary}

Its proof is given at the end of section 4.

We remark that an extension of Proposition \ref{p1} (and hence of Theorem \ref{t1}, using the same method of proof)
to the $N$-particle Jansen-Hess operator ($2<N\leq Z$) implies a critical potential strength which decreases with $N\;\;(\gamma <0.285$ for $N=Z$, due
to the relative boundedness requirement \cite[Appendix B]{Jaku2}).

\section{Proof of Proposition \ref{p1}}
\setcounter{equation}{0}

We show first that $b_{m,\theta}^{(2)}=T_\theta^{(2)}+b_{1m,\theta}^{(2)} +b_{2m,\theta}^{(2)}$ is sectorial. According to \cite{Simon} this is the case if there exists a vertex $z_0 \in {\Bbb C}$, a direction $\beta \in [0,2 \pi)$ and an opening angle $\phi \in [0,\pi)$ such that
\begin{equation}\label{4.1}
(\psi, b_{m,\theta}^{(2)}\,\psi) \subset \;\{z \in {\Bbb C}:\;|\mbox{arg}\,(e^{-i\beta}(z-z_0))|\,\leq\frac{\phi}{2}\}
\end{equation}
for $\psi \in \cA(H_{1/2}({\Bbb R}^3)\otimes {\Bbb C}^2)^2$ with $\|\psi\|=1.$

Clearly, $T_\theta^{(2)}$ is sectorial for  $\theta =e^\xi \in \cD$ because it is given by the set $\{(p\,e^{-2i \,\mbox{{\scriptsize Im}}\,\xi} +m^2)^\frac12:\,p \in {\Bbb R}_+\}$
which lies in the sector defined by $z_0=0,\;\;\beta=0$ and $\phi=2|$Im$\,\xi|\,\leq 2\xi_0.$

The $|T_\theta^{(2)}|$-form boundedness of the potential part of $b_{m,\theta}^{(2)}$ was proven in the following form
(with $\varphi:=\psi_{\bfx_1}(\bfx_2)$; see section 2),
$$|(\varphi,(b_{1m,\theta}^{(2)}+b_{2m,\theta}^{(2)})\;\varphi)|\;\leq\; \frac{1}{|\theta|}\;(\varphi,(b_1^{(2)}+b_2^{(2)})\;\varphi)\;+\;C\;(\varphi,\varphi)$$
\begin{equation}\label{4.2}
\leq\; \frac{1}{|\theta|}\;c_0\;(\varphi,p_2\,\varphi)\;+\;C\;(\varphi,\varphi),
\end{equation}
where $\frac{1}{|\theta|} \leq 1+2\xi_0,$ and $c_0=\frac{\gamma}{\gamma_{BR}}-d \gamma^2 <1$ if $\gamma<1.006.$ In turn, from (\ref{2.7}), $ \frac{1}{|\theta|} (\varphi, p_2 \varphi)\,\leq (1-\xi_0)^{-1}|(\varphi, T_\theta^{(2)} \varphi)|.$
Moreover, using  estimates similar to (\ref{2.9a}) in (\ref{4.2}), we even obtain (for $\xi_0 <\frac{\pi}{4}$)
$$|(\varphi,(b_{1m,\theta}^{(2)} +b_{2m,\theta}^{(2)})\;\varphi)|\;\leq\;c_1\;\mbox{Re }(\varphi,  T_\theta^{(2)}\,\varphi)\;+\;C\;(\varphi,\varphi)$$
\begin{equation}\label{4.3}
c_1:=\; \frac{c_0}{1-\xi_0}.
\end{equation}
Since $c_0<1$ we have $c_1<1$ for sufficiently small $\xi_0.$
According to \cite[Thm 1.33, p.320]{Kato} (\ref{4.3}) guarantees that $b_{m,\theta}^{(2)}$ as form sum is also sectorial, with the opening angle $\phi$ given by
\begin{equation}\label{4.4}
0\;<\;\tan \frac{\phi}{2}\;=\;\frac{\tan |\mbox{Im }\xi|\,+c_1}{1-c_1}\;<\infty,
\end{equation}
and some vertex $z_0<0$ which has to be sufficiently small (one has the estimate \cite[eq.(VI--1.47)]{Kato} Re $(\varphi, b_{m,\theta}^{(2)} \varphi) \,\geq -C(\varphi,\varphi)$ with the constant $C$ from (\ref{4.3})).

In the next step we prove that the spectrum $\sigma(T_\theta + a_{1,\theta})\,=
\sigma(T_\theta^{(1)} + b_{m,\theta}^{(2)})$ can be decomposed into the spectra of the two single-particle operators  according to (\ref{3.3}).

As we have just shown, $T_\theta^{(1)}$ is sectorial with maximum opening angle $\phi=2\xi_0$ and $b_{m,\theta}^{(2)}$ is sectorial  with maximum opening angle $\phi_0=:\phi(\xi_0)$ (obtained upon replacing $|$Im $\xi|$ by $\xi_0$
in (\ref{4.4}) since $\tan $ and $\arctan$ are monotonically increasing functions).
Let us take $\xi_0<\frac12$ such that $\phi +\phi_0 <\pi.\;$
This is done in the following way. Choose some $\xi_0.$ If $2\xi_0+\phi_0<\pi$, we are done.
If not, since $0<\phi_0<\pi$ there is $\delta>0$ such that $\phi_0<\delta<\pi.$
Then define $\xi_1:=\frac12( \pi-\delta)<\xi_0.$ From (\ref{4.4}) and the monotonicity of $\tan$ and $\arctan$ we have $\phi_0 >\phi(\xi_1)$ and thus
$2\xi_1 +\phi(\xi_1) <\pi.$

Writing $\psi \in \cA(H_{1/2}({\Bbb R}^3)\otimes {\Bbb C}^2)^2$ in the form domain of $T_\theta+a_{1,\theta}$ as a finite linear combination of product states $\varphi^{(1)} \varphi^{(2)}$ with $\varphi^{(k)}$ relating to particle $k$, we have
$$(\varphi^{(1)} \varphi^{(2)},(T_\theta +a_{1,\theta})\;\varphi^{(1)}\varphi^{(2)})\;=\;(\varphi^{(1)},T_\theta^{(1)}\,\varphi^{(1)})(\varphi^{(2)},\varphi^{(2)})\;
$$
\begin{equation}\label{4.5}
+\;(\varphi^{(2)},b_{m,\theta}^{(2)}\,\varphi^{(2)})\;(\varphi^{(1)},\varphi^{(1)}).
\end{equation}
Thus, the necessary assumptions for Proposition 4 of \cite{Simon} (which is based on a lemma of Ichinose) are satisfied, which guarantees
that $T_\theta +a_{1,\theta}$ is sectorial, as well as the validity of (\ref{3.3}).

\vspace{0.2cm}

Before considering the proof of the HVZ theorem for nonsymmetric potentials
we need to establish that $h_\theta^{(2)}$ as well as $T_\theta + a_{1,\theta}=:h_{0,\theta}$ is for $\theta \in \cD$ a dilation analytic family in the operator sense.
This requires in addition the relative boundedness of the potentials of $h_\theta^{(2)}$ and $h_{0,\theta}$ with respect to $T_\theta$ and $T_0$ with bound smaller
than one (such that these operators are well-defined with domain $D(T_0)$).

First we estimate with the help of (\ref{2.9a}),
\begin{equation}\label{4.5a}
\|\theta\;T_\theta\,\psi\|^2\;=\;\|(\sqrt{p_1^2+m^2\theta^2}\,+\sqrt{p_2^2+m^2\theta^2}\,)\;\psi\|^2
\end{equation}
$$\geq \;(\psi, (\mbox{Re }\sqrt{p_1^2+m^2\theta^2}\,+\,\mbox{Re }\sqrt{p_2^2+m^2\theta^2}\,)^2\;\psi)\;\geq\;
(1-\xi_0)^2\;(\psi,(p_1+p_2)^2\,\psi),$$
such that $\|T_0\psi\|\,\leq\, \frac{|\theta|}{1-\xi_0}\,\|T_\theta\psi\|.\;$
Next we decompose for $h_\theta^{(2)}=T_\theta+W_\theta$ analogously to
(\ref{2.6}),
\begin{equation}\label{4.5b}
\|W_\theta\,\psi\|\;\leq\; \frac{1}{|\theta|}\;\|W_0\psi\|\;+\;\|(W_\theta \,-\,\frac{1}{\theta}\,W_0)\;\psi\|.
\end{equation}
The boundedness of the second term in (\ref{4.5b}) follows immediately from the method of proof
of the form boundedness of $W_\theta -\frac{1}{\theta}W_0\;$ (see e.g. Appendix A).
For the first term we estimate, using $\|p_1\psi\|^2\,=\frac12 (\psi, (p_1^2+p_2^2)\,\psi)\,\leq \frac12 (\psi,(p_1+p_2)^2\,\psi),$
\begin{equation}\label{4.5c}
\|W_0\,\psi\|\;\leq\; \|\sum_{k=1}^2 (b_1^{(k)}+b_2^{(k)})\,\psi\|\;+\;\|v_0^{(12)}\psi\|\;+\;\|c_0^{(12)}\psi\|
\end{equation}
$$\leq\;\sqrt{c_w}\;\|T_0\psi\|\;+\;\frac{1}{\sqrt{2}}\;\sqrt{c_v}\,\|T_0\psi\|\;+\;2\;\frac{1}{\sqrt{2}}\,\sqrt{c_s}\;\|T_0\psi\|\;=:\tilde{c}_1\;\|T_0\psi\|$$
where $c_v=4e^4, \;\;c_w=(\frac{4}{3}\gamma +\frac{2}{9} \gamma^2)^2$ and $c_s=(\frac{2\gamma}{\pi}\,[\pi^2/4 -1])^2\,c_v $ are calculated in \cite[p.72]{thesis}.
We have $\tilde{c}_1<1$ for $\gamma \leq 0.66.$
In the same way, $\|a_{1,m=0}\psi\|\,=\,\|(b_1^{(2)}+b_2^{(2)})\,\psi\|\,\leq\, \sqrt{c_w}\,\|T_0^{(2)}\psi\|\,<0.977 \,\frac{1}{\sqrt{2}}\,\|T_0\psi\|\;$ if $\gamma \leq 0.66.\;$
With  the inequality below (\ref{4.5a})  this guarantees the relative $T_\theta$-boundedness of $W_\theta$ as well
as of $a_{1,\theta}\;$ (with bound $<1$) for $\gamma \leq 0.66$ and sufficiently small $\xi_0.$

The proof of the HVZ theorem (\ref{3.2}) is usually done in two steps.

\noindent{\it a) The easy part:} $\quad \sigma(T_\theta+a_{1,\theta})\subset \sigma_{ess}(h_\theta^{(2)})$

The proof is performed with the help of defining sequences as done in the Schr\"{o}dinger case \cite{BC} and in the $\theta =1$ Jansen-Hess case \cite{Jaku2}.
Let $\lambda \in \sigma(T_\theta +a_{1,\theta}).$
Then there exists a defining sequence $(\psi_n)_\nin$ with $\psi_n \in \cA(C_0^\infty ({\Bbb R}^3) \otimes {\Bbb C}^2)^2,\;\;\|\psi_n\|=1$ and
\begin{equation}\label{4.6}
\|(T_\theta +a_{1,\theta}-\lambda)\;\psi_n\|\;\longrightarrow \;0\qquad \mbox{ for } \ninfi.
\end{equation}
We define a unitary translation operator $T_\bfa$ by $T_\bfa \psi_n(\bfx_1,\bfx_2)=\psi_n(\bfx_1-\bfa,\bfx_2).\;$

Let $\psi_n^{(a)}:= T_\bfa\psi_n.\;$  We claim that the antisymmetrized function $\cA \psi_n^{(a)} $ is a defining
sequence for $\lambda \in \sigma(h_\theta^{(2)}).\;$
It was shown in \cite{Jaku2} that it is sufficient to prove that $\psi_n^{(a)}$ has this property.
 Since $T_\bfa$ is unitary, $\psi_n^{(a)}$ is normalized. We have

\begin{equation}\label{4.7}
\|(h_\theta^{(2)}-\lambda)\;\psi_n^{(a)}\|\;\leq\; \|(T_\theta +a_{1,\theta}-\lambda)\;\psi_n^{(a)}\|\;+\;\|r_{1,\theta}\,\psi_n^{(a)}\|.
\end{equation}
Since the only action of $T_\bfa$ is a shift of the coordinate of particle 1, it follows that $T_\bfa$ commutes with $T_\theta^{(1)}$ as well as with $b_{m,\theta}^{(2)}.$
Therefore, from (\ref{4.6}),
\begin{equation}\label{4.8}
\|(T_\theta +a_{1,\theta}-\lambda)\;T_\bfa \psi_n\|\;=\;\|T_\bfa\;(T_\theta +a_{1,\theta}-\lambda)\;\psi_n\|\;\leq\;\|T_\bfa\|\cdot \tilde{\epsilon}\;=\;\tilde{\epsilon}
\end{equation}
for a given $\tilde{\epsilon} >0$ and $n$ sufficiently large.
The second contribution to (\ref{4.7}) is decomposed into
\begin{equation}\label{4.9}
\|r_{1,\theta}\,\psi_n^{(a)}\|\;\leq\; \|b_{1m,\theta}^{(1)}\psi_n^{(a)}\|\;+\;\|b_{2m,\theta}^{(1)} \psi_n^{(a)}\|\;
+\;\|v_\theta^{(12)}\psi_n^{(a)}\|\;+\;\|c_\theta^{(12)}\psi_n^{(a)}\|.
\end{equation}
We show that the r.h.s. of (\ref{4.9}) can be made smaller that $\epsilon$ for $a$ sufficiently large.

According to Lemma 5 of \cite{Jaku2,Jaku2a} we have for $\theta=1,\;\;l=1,2$ and $\psi_n$ a finite linear combination of states $\varphi_n^{(1)} \varphi_n^{(2)}\in (C_0^\infty ({\Bbb R}^3) \otimes {\Bbb C}^2)^2$,
\begin{equation}\label{4.10}
\|b_{lm,\theta}^{(1)}\,T_\bfa\,\varphi_n^{(1)}\varphi_n^{(2)}\|\;=\;\|\varphi_n^{(2)}\|\;\|b_{lm,\theta}^{(1)}\,T_\bfa\varphi_n^{(1)}\|\;\leq\;\frac{2c}{a} \;\|\varphi_n^{(2)}\|\;\|\varphi_n^{(1)}\|
\end{equation}
with some constant $c$.
The proof of this lemma is based on the structure $W_1\frac{1}{x_1}B_1(\bfp_1)\;$ (respectively sums of such terms and their adjoints) of both $b_{1m}^{(1)}$ and $b_{2m}^{(1)}$ where $W_1$ stands for a bounded operator and $B_1(\bfp_1)$ for an analytic bounded multiplication operator in momentum space.
With the scaling property (\ref{2.3}), $\;b_{lm,\theta}^{(1)}=\frac{1}{\theta}b_{l,m\cdot \theta}^{(1)}$, and the estimates (\ref{2.5}), boundedness holds also for the dilated operators (while analyticity in $\bfp_1$ is not
affected since $\theta \neq 0).\;$
Therefore, (\ref{4.10}) holds  for all $\theta \in \cD.$

For the two-particle potentials in (\ref{4.9}) we have to proceed according to Lemma 6 of \cite{Jaku2},

\begin{lemma}\label{l1}
Let $\psi_n$ be a finite linear combination of $\varphi_n^{(1)}\varphi_n^{(2)} \in (C_0^\infty({\Bbb R}^3)\otimes {\Bbb C}^2)^2$ and  $T_\bfa$ the translation of $\bfx_1$ by $\bfa$.
Then for all $\psi \in (C_0^\infty({\Bbb R}^3)\otimes {\Bbb C}^2)^2$ and $a>4R$,
\begin{equation}\label{4.11}
|(\psi, v_\theta^{(12)}\,T_\bfa\varphi_n^{(1)}\varphi_n^{(2)})|\;\leq\;\frac{c}{a-2R}\;\|\psi\|\;\|\varphi_n^{(1)}\varphi_n^{(2)}\|
\end{equation}
\begin{equation}\label{4.12}
|(\psi,c_\theta^{(12)}\,T_\bfa \varphi_n^{(1)}\varphi_n^{(2)})|\;\leq\; \frac{c}{a-2R}\;\|\psi\|\;\|\varphi_n^{(1)}\varphi_n^{(2)}\|
\end{equation}
with some positive constants $c$ and $R$.
\end{lemma}

\begin{proof}
Take first $\theta=1$ and let $\varphi_n^{(2)}\in B_{R_2}(0)$ and $T_\bfa\varphi_n^{(1)} \in B_{R_1}(\bfa)$
where $B_R(\bfx)$ is a ball of radius $R$ centred at $\bfx$. So the inter-particle separation can be estimated by $|\bfx_1-\bfx_2|\,\geq x_1-x_2\geq a-R_1-R_2.\;$
Let $R:= \max \{R_1,R_2\}$ and $\tilde{a}:=a-2R.$ We define the smooth auxiliary function
$\chi_{12}$ mapping to $[0,1]$ by
\begin{equation}\label{4.13}
\chi_{12}\left( \frac{\bfx_1-\bfx_2}{\tilde{a}}\right):=\;\left\{ \begin{array}{cc}
0,& |\bfx_1-\bfx_2|<\tilde{a}/2\\
1,& |\bfx_1-\bfx_2|\geq \tilde{a}
\end{array}
\right.\;.
\end{equation}
Then $\chi_{12}$ is unity on the support of $T_\bfa \varphi_n^{(1)} \varphi_n^{(2)}=: \tilde{\psi},$ i.e. $\chi_{12}\tilde{\psi}=\tilde{\psi}.$

The structure of $v^{(12)}$ as well as $c^{(12)}$ is determined by terms (respectively their adjoints) of the form $W_{12}\frac{1}{|\bfx_1-\bfx_2|}\,B_1(\bfp_1) B_1(\bfp_2)$ where $W_{12}$ is a bounded (two-particle)
operator and $B_1(\bfp_k)$ are bounded analytic single-particle multiplication operators in momentum space (see (\ref{1.2}), (\ref{1.5})).
We make the decomposition (abbreviating $\tilde{B}:= B_1(\bfp_1)B_1(\bfp_2)$)
$$|(\psi, W_{12}\;\frac{1}{|\bfx_1-\bfx_2|}\,\tilde{B}\;\tilde{\psi})|\;\leq\;|(W_{12}^\ast \psi,\;\frac{1}{|\bfx_1-\bfx_2|}\,\chi_{12}\,\tilde{B}\,\tilde{\psi})|$$
\begin{equation}\label{4.14}
+\;|(W_{12}^\ast\,\psi,\;\frac{1}{|\bfx_1-\bfx_2|}\;[\chi_{12},\tilde{B}]\;\tilde{\psi})|.
\end{equation}
The first term is estimated, according to (\ref{4.13}),  by $\|W_{12}^\ast\|\;\|\psi\|\,\frac{2}{\tilde{a}}\,\|\tilde{B}\|\,\|\tilde{\psi}\|\,=\frac{c}{\tilde{a}}\,\|\psi\|\,\|\tilde{\psi}\|\;$
with some constant $c$.
The second term has already been dealt with in previous work \cite{Jaku2} by showing 
the boundedness of $\frac{1}{|\bfx_1-\bfx_2|p_k},$ as well as by proving
the uniform boundedness (with bound $c/\tilde{a})$ of the commutator $p_k[\chi_{12},\tilde{B}]\,=-p_k\,[\chi_{12,0},B_1(\bfp_1)]\,B_1(\bfp_2)\,-B_1(\bfp_1)p_k\,[\chi_{12,0},B_1(\bfp_2)]\;$ for $k \in \{1,2\}.$
The proof is done in Fourier space by profiting from the fact that $\chi_{12,0}:=1-\chi_{12}$ is a Schwartz function, and subsequently by estimating with the Lieb and Yau formula (\ref{5.8}).
\end{proof}

The proof of Lemma \ref{l1} for $\theta=1$ is easily extended to $\theta \in \cD$ with the same argumentation as given below (\ref{4.10}),
which shows that $\|v_\theta^{(12)}\psi_n^{(a)}\|\,+\,\|c_\theta^{(12)}\psi_n^{(a)}\|\,\leq \frac{\tilde{c}}{a-2R}\;$ with some constant $\tilde{c}.$

Collecting results, the r.h.s. of (\ref{4.7}) can be made arbitrarily small,
$\|(h_\theta^{(2)}-\lambda)\psi_n^{(a)}\|\,<\epsilon,$ for $a$ and $n$ sufficiently large.
This proves that $\psi_n^{(a)}$ is the required defining sequence and hence $\lambda \in \sigma(h_\theta^{(2)}).$
Since $\sigma(T_\theta+a_{1,\theta})$ is continuous according to (\ref{3.5}), we have proven
$\sigma(T_\theta+a_{1,\theta}) \subset \sigma_{ess}(h_\theta^{(2)}).$
\vspace{0.5cm}

\noindent{\it b) The hard part}:
$\quad \sigma_{ess}(h_\theta^{(2)}) \subset \sigma(T_\theta+a_{1,\theta})$

It is sufficient to prove that the spectrum of $h_\theta^{(2)}$ is discrete outside $\sigma(h_{0,\theta}).$
 We recall that $h_{0,\theta}=T_\theta +a_{1,\theta}$ and $h_\theta^{(2)}=h_{0,\theta}+r_{1,\theta}$
are well-defined operators, and from Appendix B we know that $r_{1,\theta}\,(h_{0,\theta}-z)^{-1}$ is a bounded operator for $z$ in the resolvent set of $h_{0,\theta}.$
Assuming for the moment that there is a domain in ${\Bbb C} \backslash \sigma(h_{0,\theta})$ where $r_{1,\theta}(h_{0,\theta}-z)^{-1} \neq -1\;$ (this
will follow from the implicit mapping theorem, see below), 
we get from the second resolvent identity the representation \cite{AC}
\begin{equation}\label{4.15}
\frac{1}{h_\theta^{(2)}-z}\;=\;\frac{1}{h_{0,\theta}-z}\;\frac{1}{1+r_{1,\theta}\,\frac{1}{h_{0,\theta}-z}}.
\end{equation}
We claim that the r.h.s. of (\ref{4.15}) can be extended to a meromorphic operator on ${\Bbb C}\backslash \sigma(h_{0,\theta})$ for a fixed $\theta \in \cD \backslash {\Bbb R}.$

We define the set
\begin{equation}\label{4.16}
\tilde{E}:= \;\{\sigma(h_{0,\theta}):\;\theta \in \overline{D}\}.
\end{equation}
This set is closed because $\sigma(h_{0,\theta})=\sigma_{ess}(h_{0,\theta})$ is closed. Let $\cO_z:={\Bbb C}\backslash \tilde{E}. \;\;\cO_z$ is open, nonempty and connected because $h_{0,\theta}$ is sectorial and a dilation analytic operator  in $\cD$.
Then the function
\begin{equation}\label{4.17}
F_\psi(z,\theta):= \;(\psi,r_{1,\theta}\;\frac{1}{h_{0,\theta}-z}\;\psi)
\end{equation}
for $\psi \in \cA(N_{\xi_0} \otimes {\Bbb C}^2)^2$ with $\|\psi\|=1$ is analytic in $\cD$ for all $z \in \cO_z$ since $h_{0,\theta}$ and $r_{1,\theta}$ are dilation analytic operators. Likewise, for $\theta \in \cD$ fixed, $F_\psi(z,\theta)$
is analytic in $\cO_z$ since $z$ is in the resolvent set of $h_{0,\theta}$ for all $\theta \in \cD.$
Trivially, $F_\psi(z,\theta)$ is continuous in $\cO_z \times \cD.$
Then Osgood's lemma \cite[p.3]{Gun} states that $F_\psi(z,\theta)$ is an
analytic (holomorphic) function in $\cO_z \times \cD.$

We claim that for $\theta$ fixed, $F_\psi(z,\theta) =-1$ only in isolated points of $\cO_z$, and that the multiplicity of these points is finite.

From the validity of the HVZ theorem for the Jansen-Hess operator in the case of $\theta =1\;$ (and consequently for $\theta \in \cD \cap {\Bbb R}$) which was proven  for potential strength $\gamma \leq 0.66 \;$
(this bound also results from the condition $\tilde{c}_1<1$ in (\ref{4.5c}))
we know that 
$\sigma_{ess}(h^{(2)})=\sigma(h_{0})$. This means that $(\psi, \frac{1}{h^{(2)}-z}\,\psi)\;$
and hence also $(\psi, \frac{1}{h_0-z}(1+r_1\frac{1}{h_0-z})^{-1}\psi)$ has only isolated poles (of finite multiplicity)
for $z \notin \sigma(h_{0})$
and arbitrary $\psi \in \cA(N_{\xi_0} \otimes {\Bbb C}^2)^2$. Consequently,
from the boundedness of $(h_0-z)^{-1},\;\; F_\psi(z,1)=-1$ exactly at these poles.

Let $\tilde{z} \in \sigma_d(h^{(2)})$ be one of these poles, i.e. $F_\psi(\tilde{z},1)=-1.\;$ Then $\tilde{z} \in \sigma_d(h_\theta^{(2)})$ for $\theta \in \cD \cap {\Bbb R}$. 
Therefore, due to the analyticity of $F_\psi$,  the identity theorem gives
$F_\psi(\tilde{z},\theta)=-1 $ on $\cD$.

Consider now a fixed $\tilde{\theta} \in \cD \backslash {\Bbb R} $ and choose $z_0 \in \cO_z.$ Then we claim that $F_\psi(z_0,\tilde{\theta})=-1$ implies that there is a  neighbourhood $U_{z_0}$ of $z_0$ 
such that $F_\psi({z},\tilde{\theta})\neq -1\;$ for all $z \in U_{z_0} \backslash \{z_0\}.$

We make use of the implicit mapping theorem \cite[p.19]{Gun} stating that for an analytic function $f= F_\psi+1: \cO_z \times \cD \rightarrow {\Bbb C}$
with the properties {\it (i)} $ f(z_0,\tilde{\theta})=0$ for a point $(z_0,\tilde{\theta}) \in \cO_z \times \cD$ and {\it (ii)} $ \frac{\partial f}{\partial z}(z_0,\tilde{\theta}) \neq 0$ there
exists a neighbourhood $U_{z_0} \times U_{\tilde{\theta}}$ of $(z_0,\tilde{\theta})$ such that $\forall \;\theta \in U_{\tilde{\theta}} \;\exists_1\;g(\theta) \in U_{z_0},\;\,g$ analytic:$\,\;f(g(\theta),\theta)=0.\;$
In other words, $z$ is a function of $\theta$ in $U_{\tilde{\theta}}$ which does not permit $F_\psi(z,\tilde{\theta})=-1$ in $U_{z_0}\backslash \{z_0\}.$

The proof of property {\it (ii)} is straightforward. Dealing with analytic functions, we have
\begin{equation}\label{4.18}
\frac{\partial F_\psi}{\partial z}\;=\;(\psi,r_{1,\theta}\;\frac{1}{(h_{0,\theta}-z)^2}\;\psi).
\end{equation}
Assume $\frac{\partial F_\psi}{\partial z} =0$ for the point $(z_0,\tilde{\theta}).$
Then also $F_\psi=(\psi, r_{1,\theta} \,\frac{1}{h_{0,\theta}-z}\,\psi)=0$ in $(z_0,\tilde{\theta})$ which contradicts $F_\psi(z_0,\tilde{\theta})=-1.$

Thus we have shown that for every $\theta \in \cD,\;\;F_\psi(z,\theta)=-1$ only for discrete values of $z \in \cO_z.$

Since for $\theta \in {\Bbb R}\, \cap \cD$ each of these $z$-values has finite multiplicity, analyticity of $F_\psi$ assures finite multiplicity for all $\theta \in \cD.$
We note that any accumulation point of such $z$-values could have infinite multiplicity.
However, accumulation points can only occur in $\partial \cO_z \not \subset \cO_z$ since $\cO_z$ is an open set.
Therefore, $G_\psi(z,\theta):= (\psi, \frac{1}{h_{0,\theta}-z}\,\frac{1}{1+r_{1,\theta}(h_{0,\theta}-z)^{-1}}\,\psi)\;$ and thus also $(\psi, \frac{1}{h_\theta^{(2)}-z}\,\psi)\;$
is meromorphic in $\cO_z \times \cD.$

In the last step we have to show that for every $\theta  \in \cD$ there is a meromorphic extension of $G_\psi(\cdot,\theta)$ on the open set ${\Bbb C}\backslash \sigma(h_{0,\theta}).$

Without restriction, fix $\theta \in \cD \backslash {\Bbb R}.$
Let $\tilde{F}_\psi(z,\theta):= (\psi, r_{1,\theta}\,\frac{1}{h_{0,\theta}-z}\,\psi)$ for $z \in {\Bbb C}\backslash \sigma(h_{0,\theta}).\;\;\tilde{F}_\psi(z,\theta)$
is analytic in $z$ since $z\notin \sigma(h_{0,\theta}).$
From the construction, we have $\tilde{F}_\psi(z,\theta)=F_\psi(z,\theta) $ in $ \cO_z \subset {\Bbb C}\backslash \sigma(h_{0,\theta}).$ Therefore, $\tilde{F}_\psi(\cdot, \theta)$ is the extension of $F_\psi(\cdot,\theta) $ in ${\Bbb C} \backslash \sigma(h_{0,\theta}).$

It is straightforward to show that $\tilde{G}_\psi(z,\theta):= (\psi, \frac{1}{h_{0,\theta}-z}\,\frac{1}{1+r_{1,\theta}(h_{0,\theta}-z)^{-1}}\,\psi)$ is meromorphic in ${\Bbb C} \backslash \sigma(h_{0,\theta}).$

Assume that $\tilde{F}_\psi(z,\theta)=-1$ on a subset of ${\Bbb C} \backslash \sigma(h_{0,\theta})$ which has an accumulation point.
(According to the meromorphy of $\tilde{G}_\psi \upharpoonright \cO_z=
G_\psi$ in $\cO_z$ this subset is contained
in $\tilde{E}\backslash \sigma(h_{0,\theta}).)\;$
Then $\tilde{F}_\psi(z,\theta)=-1$ on ${\Bbb C} \backslash \sigma(h_{0,\theta}).$
Since $\tilde{F}_\psi \upharpoonright \cO_z=F_\psi,$ this contradicts the fact
that $F_\psi(z,\theta) =-1$ only in isolated points.

Assume that one of those points where $\tilde{F}_\psi(z,\theta)=-1$  has infinite multiplicity.  Then according to  \cite[p.182]{Rem}, $\tilde{F}_\psi(z,\theta)=-1$ on the domain ${\Bbb C} \backslash \sigma(h_{0,\theta})$ of analyticity,
again a contradiction.

With the meromorphy of $\tilde{G}_\psi(\cdot,\theta)$ in ${\Bbb C} \backslash \sigma(h_{0,\theta})$ we have proven the meromorphy of $(\psi, \frac{1}{h_\theta^{(2)}-z}\,\psi)$ in ${\Bbb C}\backslash \sigma(h_{0,\theta})$ and hence the fact that $\sigma_{ess}(h_\theta^{(2)})$ has to be a subset of $\sigma(h_{0,\theta}).$

This completes the proof of Proposition \ref{p1}.

We note that $(h_\theta^{(2)}-z)^{-1}$ can obviously not be represented, instead of (\ref{4.15}),
in a way to be used in the form sense. This would require the introduction
of the square root of $h_{0,\theta}-z$ which is ill defined for complex $\theta$.

Concerning the proof of Corollary \ref{c1}, the improved bound on $\gamma$ (as compared to Theorem \ref{t1}) is due to the fact that the potential of $h_\theta^{BR}$ is $T_\theta$-bounded, with bound smaller than one, for $\gamma < 0.74.\;$ 
This follows from the $m=0$ estimates $\|b_1^{(k)}\psi\|^2\,\leq (\frac{4}{3}\gamma)^2\,\|p_1\psi\|^2\;$ and $|(\psi, b_1^{(1)} b_1^{(2)}\psi)|\,\leq (\frac{\gamma}{\gamma_{BR}})^2\,(\psi, p_1p_2\psi)\;$
which are obtained in Mellin space by the methods of \cite{BSS} (see also \cite[Appendix C]{thesis}). Together with a (\ref{4.5b})-type inequality  one has therefore
$$\|(b_{1m,\theta}^{(1)} + b_{1m,\theta}^{(2)})\,\psi\|\;\leq\;\frac{1}{|\theta|}\;\|(b_1^{(1)}+b_1^{(2)})\,\psi\|\;+\;\sum_{k=1}^2 \|(b_{1m,\theta}^{(k)}\,-\frac{1}{\theta}\,b_1^{(k)})\,\psi\|$$
\begin{equation}\label{4.22}
\leq\; \frac{4}{3}\,\gamma\;\frac{1}{|\theta|}\;\|T_0\psi\|\;+\;C\;\|\psi\|,
\end{equation}
such that from (\ref{4.5c}),
\begin{equation}\label{4.23}
\|(b_{1m,\theta}^{(1)}+b_{1m,\theta}^{(2)}+v^{(12})\,\psi\|\;\leq\;
\left( \frac{4}{3}\,\gamma\,+\sqrt{2}\,e^2\right)\;\frac{1}{1-\xi_0}\;\|T_\theta\psi\|\;+\;\tilde{C}\;\|\psi\|.
\end{equation}
Since the $|T_\theta^{(2)}|$-form boundedness of $b_{1m,\theta}^{(2)}\;$ with relative bound $<1$ (assuring sectoriality of $T_\theta^{(2)}+b_{1m,\theta}^{(2)})\;$ as well as the HVZ theorem for $h^{BR}$ both hold for $\gamma < \gamma_{BR}$ \cite{Tix1,BuE,Jaku2a}, these two properties cause no further restriction on the bound for $\gamma$ (as $0.74 <\gamma_{BR}).$

\section{Absence of embedded eigenvalues}
\setcounter{equation}{0}

The virial theorem for the one-particle case \cite{BE} is easily generalized to two-particle operators obeying the scaling properties (\ref{2.3}). 
Assuming that $\psi$ is an eigenfunction of $h^{(2)}$ to  some eigenvalue $\lambda$ and that $\theta \in \cD \cap {\Bbb R}_+,$ the virial theorem reads
\begin{equation}\label{5.1}
\lim_{\theta \rightarrow 1} \;(\psi_\theta,\frac{h^{(2)}(m\cdot \theta)-h^{(2)}(m)}{\theta -1}\;\psi)\;=\;\lambda\;\|\psi\|^2,
\end{equation}
where the mass dependence of $h^{(2)}$ is indicated explicitly.
By the mean value theorem, the operator on the l.h.s. is transformed into $m\,(\frac{dh^{(2)}(m)}{dm})(m\cdot \tilde{\theta})$ for some
$\tilde{\theta}$ on the line between $1$ and $\theta$.
Since this operator can be bounded independently of $\tilde{\theta}$ (see section 2) and $\|\psi_\theta\|\,=\|\psi\|,$
the theorem of dominated convergence applies and the limit $\theta \rightarrow 1$ can be carried out.
We get,
making use of the symmetry property of $\psi$
under particle exchange,
\begin{equation}\label{5.2}
\frac{\lambda}{2m}\;\|\psi\|^2\;=\;(\psi, \frac{m}{E_{p_1}}\;\psi)\,+\,(\psi, \left( \frac{db_{1m}^{(1)}}{dm}+\frac{db_{2m}^{(1)}}{dm}\,+\frac12\,\frac{dv^{(12)}}{dm}\,+\frac12\,\frac{dc^{(12)}}{dm}\right)\;\psi).
\end{equation}
This equation has to be combined with the eigenvalue equation which we take in the following form,
\begin{equation}\label{5.3}
\lambda\;(F\psi,\psi)\;=\;(F\psi,\left( \sum_{k=1}^2 (E_{p_k}+b_{1m}^{(k)}+b_{2m}^{(k)})\,+v^{(12)}\,+c^{(12)}\right)\;\psi),
\end{equation}
$$F\,\psi:=\;c_0\;(1-\frac{m}{E_{p_1}})\;\frac{1}{E_{p_1}+E_{p_2}-m}\;\psi\;=\;F^\ast\,\psi.$$
In the single-particle case, $F$ is  taken in such a way that the negative contribution (termed $\beta_{10})$ to the
linear term $\frac{db_{1m}}{dm}$ can be eliminated \cite{BE}.
Here, only a partial compensation is possible because one cannot avoid that $F(b_{1m}^{(1)}+b_{1m}^{(2)})$ is a two-particle operator 
(which cannot be split into single-particle terms).
The symmetric  (with respect to particle exchange) energy denominator $(E_{p_1}+E_{p_2}-m)^{-1}$ assures that the
operator $Fh^{(2)}$ appearing on the r.h.s. of (\ref{5.3}) is bounded.
 $c_0 \in {\Bbb R}_+$ is a parameter to be determined later.
Let us now restrict ourselves to the Brown-Ravenhall operator. Then we have

\begin{proposition}\label{p2}
Let $h^{BR}$ be the two-particle Brown-Ravenhall operator and assume $\gamma \leq\gamma_c$ with $\gamma_c = 0.37\;\;(Z\leq 50).\;$
Then there are no eigenvalues in $[2m,\infty).$
\end{proposition}

Note that, with $\sigma_{ess}(h^{BR})=[\Sigma_0,\infty)$ and $\Sigma_0 <2m$ \cite{Mor}, 
no information on embedded eigenvalues is provided  for the subset $[\Sigma_0,2m)$.
This corresponds to the multi-particle Schr\"{o}dinger case where the 
virial-theorem method provides the absence of eigenvalues  only in the subset $[0,\infty)$ of the essential spectrum \cite[p.232]{Reed4}.

\begin{proof} 
Defining $d\bfpi:=d\bfp_1 d\bfp_2 d\bfp'_1 d\bfp'_2$ we have from \cite{BE}
\begin{equation}\label{5.5}
(\psi,\frac{db_{1m}^{(1)}}{dm}\;\psi)\;=\;\beta_{10}(m)\;+\;\beta_{11}(m),
\end{equation}
$$\beta_{10}(m):=\; \mbox{Re }(\psi, \left( \frac{1}{E_{p_1}}\,-\,\frac{m}{E_{p_1}^2}\right)\,b_{1m}^{(1)}\;\psi)$$
$$\beta_{11}(m):=\;\frac{\gamma}{2\pi^2}\int_{{\Bbb R}^{12}} d\bfpi\;\overline{g(p_1)\bfsigma^{(1)}\bfp_1\,\hat{\psi}(\bfp_1,\bfp_2)}\;\frac{1}{|\bfp_1-\bfp'_1|^2}$$
$$\cdot \left( \frac{1}{E_{p_1}}\,+\,\frac{1}{E_{p'_1}}\right)\;g(p'_1)\,\bfsigma^{(1)}\bfp'_1\;\delta(\bfp_2-\bfp'_2)\;\hat{\psi}(\bfp'_1,\bfp'_2).$$
Subtraction of the real part of   (\ref{5.3}) from (\ref{5.2}),
while dropping the second-order terms $b_{2m}^{(k)} $ and $c^{(12)}$, results in
\begin{equation}\label{5.6}
0\;=\; M_0\;+\;\gamma\,M_1\;+\;e^2\,M_2,
\end{equation}
\begin{eqnarray*}
M_0&:=& (\psi, \left( (1\,-\frac{\lambda}{2m})\,(1\;-\;\frac{2c_0m(1-\frac{m}{E_{p_1}})}{E_{p_1}+E_{p_2}-m})\right.\\
&&\left.-\;(1\,-\,\frac{m}{E_{p_1}})\,(1\;+\;\frac{c_0(E_{p_1}+E_{p_2}-2m)}{E_{p_1}+E_{p_2}-m}) \right)\;\psi)\\
\gamma\,M_1&:=& \beta_{10}(m)\;-\;\mbox{Re }(\psi, c_0(1\,-\,\frac{m}{E_{p_1}})\;\frac{1}{E_{p_1}+E_{p_2}-m}\\
&&\cdot(b_{1m}^{(1)}+b_{1m}^{(2)})\;\psi)\;+\;\beta_{11}(m)\\
e^2\,M_2&:=& \mbox{Re } (\psi,(1\,-\,\frac{m}{E_{p_1}})\,(\frac{1}{E_{p_1}}\;-\;\frac{c_0}{E_{p_1}+E_{p_2}-m})\;v^{(12)}\;\psi)\\
&&-\int_{{\Bbb R}^{12}}d\bfpi\;\overline{g(p_1)\,\bfsigma^{(1)}\bfp_1\,U_0^{(2)\ast}\,\hat{\psi}_0(\bfp_1,\bfp_2)}\;(\frac{1}{E_{p_1}}\,+\,\frac{1}{E_{p'_1}})\\
&&\cdot k_{v^{(12)}}\;\left( g(p'_1)\,\bfsigma^{(1)} \bfp'_1\,U_0^{(2')\ast}\;\hat{\psi}_0(\bfp'_1,\bfp'_2)\right).
\end{eqnarray*}
In the expression for the electron-electron interaction term, $ e^2M_2,$ it is used that $b_{1m}^{(1)}$ and $v^{(12)}$ have the same structure.
Indeed, the kernel of  $b_{1m}^{(1)}$ is given by 
 $U_0^{(1)}k_{b_{1m}}U_0^{(1')\ast}$  with
\begin{equation}\label{5.7}
k_{b_{1m}}:=\; -\frac{\gamma}{2\pi^2}\,\frac{1}{|\bfp_1-\bfp'_1|^2}\,\delta(\bfp_2-\bfp'_2),
\end{equation}
as compared to the kernel of $v^{(12)}$ defined above (\ref{2.10a}).
Due to the symmetry upon particle exchange, the kernel of $\frac12 \frac{dv^{(12)}}{dm}$ in (\ref{5.2}) 
can be replaced by $U_0^{(2)}\,\frac{d}{dm}(U_0^{(1)}k_{v^{(12)}}U_0^{(1)\ast})\,U_0^{(2)\ast}.$
 Therefore
 (\ref{5.5}), with $k_{v^{(12)}}$ substituted for  $k_{b_{1m}},$  is applicable.
As in section 2, $\psi_0={\psi \choose 0}$ is a two-particle spinor with the lower components set equal to zero.

For a symmetric integral operator $\cO$ with kernel $K+K^\ast$, we use the Lieb and Yau formula, derived from the Schwarz inequality, in the following form \cite{Lieb} (see also \cite{Jaku2a})
\begin{equation}\label{5.8}
|(\psi, \cO \,\psi)|\;\leq\; \int_{{\Bbb R}^6} d\bfp_1 d\bfp_2\;|\hat{\psi}(\bfp_1,\bfp_2)|^2\;(I_1(\bfp_1,\bfp_2)\,+\,I_2(\bfp_1,\bfp_2))
\end{equation}
$$I_1(\bfp_1,\bfp_2):=\; \int_{{\Bbb R}^6} d\bfp'_1 d\bfp'_2\;|K(\bfp_1,\bfp_2;\bfp'_1,\bfp'_2)|\;\frac{f(p_1)}{f(p'_1)}\;\frac{g(p_2)}{g(p'_2)}$$
and $I_2$ results from the replacement of $K(\bfp_1,\bfp_2;\bfp'_1,\bfp'_2)$ by $K^\ast(\bfp'_1,\bfp'_2;\bfp_1,\bfp_2).\;\;f$ and $g$ are suitable nonnegative convergence generating functions such that
$I_1,\;I_2$ exist as bounded functions for $\bfp_1,\bfp_2 \in {\Bbb R}^3.\;$
In order to get rid of the particle mass $m$, we introduce the new variables
$\bfp_i=:m\bfq_i,\;\;\bfp'_i=:m\bfq'_i,\;\;i=1,2.\;$
With $s:= 1-2c_0(1-\frac{1}{\sqrt{q_1^2+1}})/(\sqrt{q_1^2+1}+\sqrt{q_2^2+1}-1)\;$ we estimate
\begin{equation}\label{5.9}
0\;\leq\; M_0\,+\,\gamma\;|M_1|\;+\;e^2\;|M_2|
\end{equation}
$$\leq\; m^6 \int_{{\Bbb R}^6} d\bfq_1 d\bfq_2\;|\hat{\psi}(m\bfq_1,m\bfq_2)|^2\;s\left(1\,-\,\frac{\lambda}{2m}\,+\,\phi(q_1,q_2)\right).$$
For $c_0<2$  (or $c_0\leq 2$ if $q_2 \neq 0)\;$ we have $s>0$ and then
$$\phi(q_1,q_2):=\;\frac{1}{s}\;\{ -\,(1-\,\frac{1}{\sqrt{q_1^2+1}})\left( 1\,+\,c_0\,\frac{\sqrt{q_1^2+1}+\sqrt{q_2^2+1}-2}{\sqrt{q_1^2+1}+\sqrt{q_2^2+1}-1}\right)$$
\begin{equation}\label{5.10}
+\;\gamma\,q_1^2\,\tilde{M}_1\;+\;e^2\,q_1^2\,\tilde{M}_2\}
\end{equation}
where $q_1^2 \tilde{M}_i,\;\;i=1,2,\;$ result from the estimates of $M_i$ and are given in Appendix C.
From (\ref{5.9}) it follows that if $\phi(q_1,q_2) <0,\;$ we need $1-\frac{\lambda}{2m}>0$ 
which confines $\lambda$ to $\lambda <2m.$
A numerical investigation shows that the supremum of $s\,\phi(q_1,q_2)$ is attained for $q_1,q_2 \rightarrow \infty$ with $q_1 \ll q_2.$
Then $s \rightarrow 1$ and from the explicit expression (see Appendix C) it follows that
\begin{equation}\label{5.11}
\sup_{q_1,q_2 \geq 0}\,s\;\phi(q_1,q_2)\;=\; \lim_{{q_1 \rightarrow \infty \atop q_2\gg q_1}} \phi(q_1,q_2)\;=\;-(1+c_0)\,+\,\gamma\,(4+2c_0)\,+\,4\,e^2.
\end{equation}
For the optimum choice $c_0=2,$ we obtain $\sup\limits_{q_1,q_2 \geq 0}\,s\,\phi(q_1,q_2)=0$ for $\gamma =:\gamma_c =0.37.$
\end{proof}

The proof of Proposition \ref{p2} can readily be extended to the Jansen-Hess operator $h^{(2)}.$ However, the  so obtained critical potential strength $\gamma_c$
is expected to be rather small. Note that inclusion of the second-order term in the single-particle case leads to a reduction
from $\gamma_c = \gamma_{BR}\approx 0.906$ to $\gamma_c =0.29$ \cite{Jaku1}.

\section*{Appendix A \quad (Boundedness of $c^{(12)}(m\cdot \theta) -c_0^{(12)})$}
\renewcommand{\theequation}{\Alph{section}.\arabic{equation}}
\setcounter{equation}{0}
\setcounter{section}{1}

From (\ref{1.5}) and (\ref{1.6}) one derives for the $k=1$ contribution to the
 kernel of this operator  \cite{Jaku3}, using the mean value theorem,
$$\left| (K^{(1)}_{c^{(12)}(m\cdot \theta)}-K^{(1)}_{c_0^{(12)}})(\bfp_1,\bfp_2;\bfp'_1,\bfp'_2)\right|\;\leq\; \frac{ \gamma e^2m}{(2\pi)^4}\;\frac{1}{|\bfp_2-\bfp'_2|^2}\;\frac{1}{|\bfp_2-\bfp'_2+\bfp_1-\bfp'_1|^2}$$
\begin{equation}\label{A.1}
\cdot \left|\frac{\partial}{\partial m}\;\{ U_0^{(1)}U_0^{(2)}\;[
\frac{1}{E_{|\bfp_2-\bfp'_2+\bfp_1|}+E_{p'_1}}\left( 1+\tilde{D}_0^{(1)}(\bfp'_1)-\tilde{D}_0^{(1)}(\bfp_2-\bfp'_2+\bfp_1)
\right.\right.
\end{equation}
$$\left.\left.
-\tilde{D}_0^{(1)}(\bfp_2-\bfp'_2+\bfp_1)\tilde{D}_0^{(1)}(\bfp'_1)\right)\;+\;h.c.\,]\;\;U_0^{(1')\ast}U_0^{(2')\ast}\,\}(\tilde{m}_1\cdot \theta)\;+\;(\tilde{m}_1 \mapsto \tilde{m}_2)\right|$$
where $h.c.$ denotes the hermitean conjugate of the first term together with the replacement  $(\bfp_1,\bfp_2) \leftrightarrows (\bfp'_1,\bfp'_2).\;$
(The second contribution $(k=2)$ to the kernel arises from particle exchange and is therefore bounded by the same constant.)
After carrying out the derivative, the modulus of each of the resulting terms is estimated separately, using the boundedness
of the dilated $U_0^{(k)},\;\tilde{D}_0^{(k)}$ and the estimate (\ref{2.5a}) for the dilated energy denominator.
According to the Lieb and Yau formula (\ref{5.8}),
 $c^{(12)}(m\cdot \theta)-c_0^{(12)}$ is bounded if the integral
\begin{equation}\label{A.2}
I(\bfp_1,\bfp_2):=\;\int_{{\Bbb R}^6} \!\!\!\! d\bfp'_1d\bfp'_2\,\left| (K^{(1)}_{c^{(12)}(m\cdot \theta)}-K^{(1)}_{c_0^{(12)}})(\bfp_1,\bfp_2;\bfp'_1,\bfp'_2)\right|\;\frac{f(p_1)g(p_2)}{f(p'_1)g(p'_2)}
\end{equation}
is bounded for all $\bfp_1,\bfp_2 \in {\Bbb R}^3$, where $f,g\geq 0$ are suitably chosen  functions.
The derivative of the  operator $\tilde{D}_{0,\theta}^{(k)}$ can be estimated by $\frac{c}{p_k}$  because $\tilde{D}_0^{(k)}$ is bounded and its $m$-dependence enters only via $m/p_k\;$ (see (\ref{2.7c})).
Finally, since from (\ref{2.5}),
$\left| \frac{m\theta}{E_\theta(p')}\right|\,\leq m|\theta|\,\frac{1}{(1-\xi_0)E_{p'}}\,\leq (1-\xi_0)^{-2}$,
the 
derivative of the energy denominator is estimated by $\left| \frac{\partial}{\partial m}\;\frac{1}{E_\theta(p)+E_\theta(p')}\right|\,\leq\,\frac{c}{p}\,\frac{1}{p+p'}.\;$

For reasons of convergence we have to keep, however, the $m$-dependence of the energy denominator in those contributions to (\ref{A.1})
which contain the factor $1/p'_2$ from the estimate of the derivatives.
This can be handled in the following way:
Let $f(m)g(m)-f(0)g(0)=[f(m)-f(0)]g(m)\,+f(0)[g(m)-g(0)]\;$
and interpret $g$ as the energy denominator and $f$ as the adjacent factors inside the curly bracket in (\ref{A.1}).
Then, while estimating the derivative of $f(m)$ by an $m$-independent function (in general setting $m=0$), the energy denominator can be estimated, using (\ref{2.5a}), by
$$\left| \frac{1}{E_\theta(|\bfp_2-\bfp'_2+\bfp_1|)+E_\theta(p'_1)}\right|\;\leq\; \frac{1}{(1-\xi_0)^3}\;\frac{1}{E_{|\bfp_2-\bfp'_2+\bfp_1|}+E_{p'_1}}$$
\begin{equation}\label{A.3}
\leq\;\frac{c}{|\bfp_2-\bfp'_2+\bfp_1|+p'_1+1}
\end{equation}
which relies on $m\neq 0.$

For the sake of demonstration we select the contribution to (\ref{A.1}) which contains the derivative of $U_0^{(2')\ast},$ leading to the estimate $\frac{\tilde{c}}{p'_2}$ according to (\ref{2.7c}).
Absorbing the bounds of $U_0^{(1)},\;U_0^{(2)},\;U_0^{(1')\ast}$ and $\tilde{D}_0^{(1)}$ into the generic constant $c$, we get for the respective contribution, say $\tilde{I},$ to (\ref{A.2}),
$$\tilde{I}(\bfp_1,\bfp_2)\;\leq\; m\cdot c \int_{{\Bbb R}^6}\!\!d\bfp'_1 d\bfp'_2\;\frac{1}{|\bfp_2-\bfp'_2|^2}\;\frac{1}{|\bfp_2-\bfp'_2+\bfp_1-\bfp'_1|^2}$$
\begin{equation}\label{A.4}
\cdot\frac{1}{|\bfp_2-\bfp'_2+\bfp_1|+p'_1+1}\cdot \frac{1}{p'_2}\;\frac{f(p_1)g(p_2)}{f(p'_1)g(p'_2)}.
\end{equation}
We choose $f(p)=p^\frac12$ and $g(p)=p.\;$ Making the substitution $\bfq:=\bfp'_2-\bfp_2$ for $\bfp'_2$
and defining $\bfxi_2:= \bfq-\bfp_1$ we have
\begin{equation}\label{A.5}
\tilde{I}(\bfp_1,\bfp_2)\;\leq\; m\cdot c\,p_1^\frac12\int_{{\Bbb R}^3}\!\!\!\! d\bfq \;\frac{1}{q^2}\;\frac{p_2}{|\bfq+\bfp_2|^2}\int_{{\Bbb R}^3}\!\!\!\!d\bfp'_1\;\frac{1}{|\bfxi_2+\bfp'_1|^2}\;\frac{1}{\xi_2+p'_1+1}\;\frac{1}{p_1^{'1/2}}.
\end{equation}
For $\xi_2=0$, the second integral is bounded. For $\xi_2\neq 0,$ let $y:=p'_1/\xi_2.\;$
Then the second integral turns into \cite[Appendix A]{thesis}
$$\frac{2\pi}{\xi_2}\int_0^\infty \frac{dp'_1}{\xi_2+p'_1+1}\;p_1^{'1/2}\;\ln\frac{\xi_2+p'_1}{|\xi_2-p'_1|}$$
\begin{equation}\label{A.6}
=\; 2\pi\int_0^\infty \frac{dy}{y^\frac12}\;\ln \frac{1+y}{|1-y|}\cdot \frac{\xi_2^\frac12 \,y}{\xi_2(1+y)+1}\;\leq\;\frac{\tilde{c}}{1+\xi_2^{1/2}}
\end{equation}
since the last factor can be estimated by $\frac{c}{1+\xi_2^{1/2}}$ and the remaining integral is convergent.
Therefore we get with the substitution $\bfq_2:= \bfq/p_2$,
\begin{equation}\label{A.7}
\tilde{I}(\bfp_1,\bfp_2)\;\leq\; m\;c\,\tilde{c}\int_{{\Bbb R}^3}\!\!
d\bfq_2\;\frac{1}{q_2^2}\;\frac{1}{|\bfq_2+\bfe_{p_2}|^2}\cdot \frac{1}{p_1^{-\frac12}+|\bfq_2 p_2/p_1\,-\bfe_{p_1}|^\frac12}
\end{equation}
where $\bfe_{p_i}$ is the unit vector in the direction of $\bfp_i,\;\;i=1,2.\;$
The last factor is bounded for $p_1<\infty$, and the remaining integral is finite.
For $p_1\rightarrow \infty$, one gets at most an additional square-root singularity, which is integrable. Thus $\tilde{I} $ is finite.

The contribution to (\ref{A.1}) arising from the derivative of $U_0^{(2)}$ which is estimated by $\frac{c}{p_2}$, is handled by the same integrals if one chooses $g(p)=p^2$ instead of $g(p)=p.$
For the boundedness of the remaining contributions to (\ref{A.1}) one can use similar techniques as for the proof of the $p$-form boundedness
of $c^{(12)}$ \cite{Jaku3}. One must, however, take care to use the same convergence generating functions in the corresponding
hermitean conjugate term entering into the r.h.s. of (\ref{A.1}). (For example, in the estimates of the derivative of $\tilde{D}_0^{(1)}$, one should take $f(p)=p^{\frac{3}{2}}$ and $g(p)=1.$)

\section*{Appendix B \quad (Boundedness of $r_{1,\theta}\,(h_{0,\theta}-z)^{-1}$)}
\setcounter{equation}{0}
\setcounter{section}{2}

Let $z \in {\Bbb C} \backslash \sigma(h_{0,\theta})$ such that $(h_{0,\theta}-z)^{-1}$ is bounded. From (\ref{4.5a})ff we have
\begin{equation}\label{B.1}
\|a_{1,\theta}\psi\|\;\leq\; c'_1\;\|T_\theta \psi\|\;+\;C_1\;\|\psi\|
\end{equation}
with $c'_1:= \sqrt{\frac{c_w}{2}}\,\frac{1}{1-\xi_0}\,<1$ for a suitable $\xi_0 < \frac12$ and $\gamma \leq 0.66$, and some constant $C_1$.
By the same estimates,
$$\|r_{1,\theta}\psi\|\;\leq\; \frac{1}{|\theta|}\left( \|(b_1^{(1)}+b_2^{(1)})\,\psi\|\;+\;\|v^{(12)}\psi\|\;+\;\|c^{(12)}\psi\|\right)\;+\;C_2\;\|\psi\|$$
\begin{equation}\label{B.2}
\leq\; c_2\;\|T_\theta \psi\|\;+\;C_2\;\|\psi\|
\end{equation}
with $c_2:= (\sqrt{\frac{c_w}{2}}+\sqrt{\frac{c_v}{2}} + \sqrt{2c_s})\,\frac{1}{1-\xi_0}\,<1\;$ if $\;\tilde{c}_1/(1-\xi_0)<1$
(with $\tilde{c}_1$ from (\ref{4.5c})).

Then, applying (\ref{B.2}) to $\tilde{\psi}:= (h_{0,\theta}-z)^{-1}\psi,$
\begin{equation}\label{B.3}
\|r_{1,\theta}\,\frac{1}{h_{0,\theta}-z}\,\psi\|\;\leq\; c_2\;\|T_\theta\;\frac{1}{h_{0,\theta}-z}\,\psi\|\;+\;C_2\;\|\frac{1}{h_{0,\theta}-z}\,\psi\|.
\end{equation}
The last term is bounded. With the second resolvent identity,
$(h_{0,\theta}-z)^{-1}\,=(T_\theta -z)^{-1}-(T_\theta-z)^{-1}a_{1,\theta}\,(h_{0,\theta}-z)^{-1},$
we estimate the first term,
\begin{equation}\label{B.4}
\|T_\theta\;\frac{1}{h_{0,\theta}-z}\,\psi\|\;\leq\; \|T_\theta\,\frac{1}{T_\theta-z}\,\psi\|\;+\;\|T_\theta\,\frac{1}{T_\theta -z}\|\cdot \|a_{1,\theta}\;\frac{1}{h_{0,\theta}-z}\,\psi\|.
\end{equation}
Note that $\sigma(T_\theta) \subset \sigma(h_{0,\theta})$ according to (\ref{3.5}) such that $(T_\theta -z)^{-1}$ is bounded. From 
$\|T_\theta (T_\theta-z)^{-1}\psi\|\,\leq \|\psi\|\,+\,|z|\,\|(T_\theta-z)^{-1}\psi\|$ the first term in (\ref{B.4}) is also bounded.
Finally, from (\ref{B.1}) and with $T_\theta = h_{0,\theta} -a_{1,\theta},$
$$\|a_{1,\theta}\;\frac{1}{h_{0,\theta}-z}\,\psi\|\;\leq\; c'_1\;\|T_\theta\;\frac{1}{h_{0,\theta}-z}\,\psi\|\;+\;C_1\;\|\frac{1}{h_{0,\theta}-z}\,\psi\|$$
\begin{equation}\label{B.5}
\leq \;c'_1\;\|h_{0,\theta}\;\frac{1}{h_{0,\theta}-z}\,\psi\|\;+\;c'_1\;\|a_{1,\theta}\;\frac{1}{h_{0,\theta}-z}\,\psi\|\;+\;C_1\;\|\frac{1}{h_{0,\theta}-z}\,\psi\|.
\end{equation}
Rearranging (\ref{B.5}) one gets 
\begin{equation}\label{B.6}
\|a_{1,\theta}\,\frac{1}{h_{0,\theta}-z}\,\psi\|\;\leq\; \frac{c'_1}{1-c'_1}\;\|h_{0,\theta} \,\frac{1}{h_{0,\theta}-z}\,\psi\|\;+\;\frac{C_1}{1-c'_1}\;\|\frac{1}{h_{0,\theta}-z}\,\psi\|,
\end{equation}
the r.h.s. being obviously bounded.

\section*{Appendix C \quad (Estimates for $\gamma M_1$ and $e^2M_2$)}
\setcounter{equation}{0}
\setcounter{section}{3}

From (\ref{5.5}) and (\ref{5.6}) we have
$$|\gamma M_1|\,\leq\, \frac12\left| \int_{{\Bbb R}^{12}} d\bfpi \overline{\hat{\psi}(\bfp_1,\bfp_2)}\left\{ (1\,-\,\frac{m}{E_{p_1}})\,(\frac{1}{E_{p_1}}\,-\,\frac{c_0}{E_{p_1}+E_{p_2}-m})\,b_{1m}^{(1)}
\right.\right.$$
$$+\,b_{1m}^{(1)}\,(1\,-\,\frac{m}{E_{p_1}})
\, (\frac{1}{E_{p_1}}\,-\,\frac{c_0}{E_{p_1}+E_{p_2}-m})\,-\,c_0\,(1\,-\,\frac{m}{E_{p_1}})\,\frac{1}{E_{p_1}+E_{p_2}-m}\,b_{1m}^{(2)}$$
\begin{equation}\label{C.1}
-\,c_0\,b_{1m}^{(2)}\,(1\,-\,\frac{m}{E_{p_1}})\,
 \left.\left. \frac{1}{E_{p_1}+E_{p_2}-m}\right\} \,\hat{\psi}(\bfp_1,\bfp_2)\right|\;+\;|\beta_{11}(m)|.
\end{equation}
Each of the four terms in curly brackets is estimated separately by its modulus.
For the sake of demonstration we select the second term.
With $k_{b_{1m}}$ from (\ref{5.7}) and the Lieb and Yau formula (\ref{5.8}), we get
$$T_b:=\;\frac12 \left| \int_{{\Bbb R}^{12}} d\bfpi \;\overline{\hat{\psi}_0(\bfp_1,\bfp_2)}\,U_0^{(1)}\,k_{b_{1m}}U_0^{(1')\ast}\left( 1\,-\,\frac{m}{E_{p'_1}}\right) \right.$$
\begin{equation}\label{C.2}
\left. \cdot \left( \frac{1}{E_{p'_1}}\,-\,\frac{c_0}{E_{p'_1}+E_{p'_2}-m}\right)\,\hat{\psi}_0(\bfp'_1,\bfp'_2)\right|
\end{equation}
$$\leq\; \frac12 \int_{{\Bbb R}^6} d\bfp_1 d\bfp_2 \,\left| U_0^{(1)\ast}\,\hat{\psi}_0(\bfp_1,\bfp_2)\right|^2\cdot I_b.$$
Taking $f(p_1)=\frac{p_1^{5/2}}{\sqrt{p_1^2+m^2}+m},\;\;g=1,\;$
and estimating
$|\frac{1}{E_{p'_1}}\,-\frac{c_0}{E_{p'_1}+E_{p'_2}-m}|\,\leq \frac{1}{E_{p'_1}}\;$
(which holds for $c_0 \leq 2)$ we obtain in the new variables $\bfq_i,\bfq'_i$ after performing the angular integration in
the variable $\bfq'_1$ \cite[Appendix A]{thesis},
$$I_b:=\,\frac{\gamma}{2\pi^2}\int_{{\Bbb R}^6} d\bfp'_1 d\bfp'_2\,(1\,-\,\frac{m}{E_{p'_1}})\,\left| \frac{1}{E_{p'_1}}\,-\,\frac{c_0}{E_{p'_1}+E_{p'_2}-m}\right|$$
\begin{equation}\label{C.3}
\cdot \frac{1}{|\bfp_1-\bfp'_1|^2}\;\delta(\bfp_2-\bfp'_2)\;\frac{f(p_1)}{f(p'_1)}
\end{equation}
$$\leq\; \frac{\gamma}{\pi}\;q_1^{3/2}\;\frac{1}{\sqrt{q_1^2+1}+1}\int_0^\infty dq'_1\;\ln\;\frac{q_1+q'_1}{|q_1-q'_1|}\;q_1^{'\frac12}\;\frac{1}{q^{'2}_1+1}.$$
In order to get an analytical estimate of (\ref{C.3}) we use
\begin{equation}\label{C.4}
\frac{1}{q^{'2}_1+1}\;\leq\;\left\{\begin{array}{ll}
1,&q'_1 \leq 1\\
&\\
\displaystyle\frac{1}{q^{'2}_1},& q'_1>1
\end{array} \right.
\end{equation}
such that, upon substituting $q'_1=:q_1\,z$ \cite[Appendix A]{thesis},
$$I_b\;\leq\; \frac{\gamma}{\pi}\;\frac{q_1^3}{\sqrt{q_1^2+1}+1}\left[
\int_0^{1/q_1} dz\,z^\frac12 \,\ln \frac{1+z}{|1-z|}\;+\;\frac{1}{q_1^2}\int_{1/q_1}^\infty \frac{dz}{z^{3/2}}\;\ln\,\frac{1+z}{|1-z|}\right]$$
\begin{equation}\label{C.5}
=\; \frac{\gamma}{\pi}\,q_1^2\;\frac{1}{\sqrt{q_1^2+1}+1}\,\left[ q_1F_{1/2}(\frac{1}{q_1})\,+\,\frac{1}{q_1}\,G_{-3/2}(\frac{1}{q_1})\right],
\end{equation}
$$F_{1/2}(a):=\,\frac{2}{3}\,\left[ a^{3/2} \ln\,\left|\frac{1+a}{1-a}\right|\,+\,4\sqrt{a}\,-\,2 \arctan \sqrt{a} \,-\,\ln \,\left|\frac{1+\sqrt{a}}{1-\sqrt{a}}\right|\,\right]$$
$$G_{-3/2}(a):=\,2\pi\,-\,2\ln\,\left| \frac{\sqrt{a}+1}{\sqrt{a}-1}\right|\,-\,4\arctan \sqrt{a}\,+\,\frac{2}{\sqrt{a}} \,\ln\,\left| \frac{1+a}{1-a}\right|.$$
For the first contribution to $|\gamma M_1|,$ the same functions $f,g$ have to be taken,
and the approximation $\sqrt{q_1^{'2}+1}\leq q'_1+1\,$ is made to allow for an analytic evaluation of the corresponding integral.
For the third and fourth contribution to $|\gamma M_1|$  we use instead $f=1,\;\;g(p_2)=p_2^{3/2}$ and the additional estimate (for $c\geq 0)$
\begin{equation}\label{C.6}
\frac{1}{\sqrt{q^{'2}+1}+c}\;\leq\;\left\{\begin{array}{ll}
\displaystyle\frac{1}{1+c},& q' \leq 1+c\\
&\\
\displaystyle\frac{1}{q'},& q' > 1+c
\end{array} \right.\,.
\end{equation}
For the estimate of $|\beta_{11}(m)|$ we define $\psi_1:= g(p_1)\bfsigma^{(1)}\bfp_1\,\psi,\;$ take $f(p_1)=p_1^{3/2},\;g=1\;$ and use again (\ref{C.6}).
With $|U_0^{(1)\ast}\hat{\psi}_0(\bfp_1,\bfp_2)|^2\,=|\hat{\psi}(\bfp_1,\bfp_2)|^2\,$ and
$\,|\hat{\psi}_1(\bfp_1,\bfp_2)|^2
=\,\frac{q_1^2}{2\sqrt{q_1^2+1}(\sqrt{q_1^2+1}+1)}\,|\hat{\psi}(\bfp_1,\bfp_2)|^2\;$ we then obtain
$$q_1^2\,\tilde{M}_1=\,\frac{q_1^2}{l\,(l+1)}\left\{ \,\left| \frac{1}{l}\,-\,\frac{c_0}{l\,+\sqrt{q_2^2+1}\,-1} \right|\frac{q_1(q_1+2)}{l\,+1}\right.$$
$$+\;\frac{l}{2\pi}\left( q_1F_{1/2}(\frac{1}{q_1})\,+\,\frac{1}{q_1}\,G_{-3/2}(\frac{1}{q_1})\right)+\,c_0\,\frac{q_2}{l\,+\sqrt{q_2^2+1}\,-1}$$
\begin{equation}\label{C.7}
 +\,\frac{c_0}{2\pi}\left( \frac{q_2}{l}\,F_{-1/2}(\frac{l}{q_2})\,+\,G_{-3/2}(\frac{l}{q_2})\right)
\end{equation}
$$\left. +\,\frac{1}{2\pi}\left( \frac{2\pi q_1}{l}
\,+\,q_1F_{-1/2}(\frac{1}{q_1})\,+\,G_{-3/2}(\frac{1}{q_1})\right)\right\}$$
where $l=\sqrt{q_1^2+1}$ and
\begin{equation}\label{C.8}
F_{-1/2}(a):=\;2\sqrt{a} \,\ln\,\left| \frac{1+a}{1-a}\right|\,+\,4\,\arctan \sqrt{a}\,-\,2\,\ln\left|\frac{\sqrt{a}+1}{\sqrt{a}-1}\right|.
\end{equation}
For estimating $e^2 M_2$ the same techniques are used, except for the simpler estimate $\frac{1}{\sqrt{q^{'2}_1+1}}\,\leq \frac{1}{q'_1}\,$
in the last contribution
(which has little effect on $\gamma_c$ due to the smallness of $e^2$).
This results in
$$q_1^2\tilde{M}_2\;=\;\frac{q_1^2}{l\,(l\,+1)}\left\{\,\left| \frac{1}{l}\,-\,\frac{c_0}{l\,+\sqrt{q_2^2+1}\,-1}\right|\,
\frac{q_1(q_1+2)}{l\,+1}
\right.$$
\begin{equation}\label{C.9}
\left.+\;\frac{l}{2\pi}\left( q_1F_{1/2}(\frac{1}{q_1})\,+\,\frac{1}{q_1}\;G_{-3/2}(\frac{1}{q_1})\right)\,+\,\left( \frac{q_1}{l}\,+\,1\right)\right\}.
\end{equation}

\section*{Acknowledgment}

I would like to thank L.Erd\"{o}s for valuable comments and H.Kalf, H.Siedentop and M.Huber for helpful discussions.
Partial support by the EU network Analysis and Quantum (contract HPRN-CT-2002-00277) is gratefully acknowledged.

\vspace{1cm}

\end{document}